\documentclass[aps,12pt,superscriptaddress,preprint,tightenlines,floatfix,preprintnumbers]{revtex4-1}
\usepackage{epsfig}
\usepackage{latexsym}
\usepackage{amsmath,amssymb,amsthm,graphicx,bm,color}
\usepackage{rotating} 
\usepackage{bm}
\newcommand{\cL}{{\cal L}} 
 
\newcommand{\tr}{\text{tr\,}} 

\newcommand{\eps}{\epsilon}
\newcommand{\ve}{\varepsilon}

\renewcommand{\epsilon}{\varepsilon}
\theoremstyle{plain}

\newtheorem{theorem}{Theorem}
\theoremstyle{definition}

\begin{document}

\title{The tight knot spectrum in QCD}
\author{Roman V. Buniy}
\email{roman.buniy@gmail.com}
\affiliation{Chapman University, Schmid College of Science, Orange, CA 92866}
\affiliation{Isaac Newton Institute, University of Cambridge, Cambridge, CB3 0EH, UK}
\author{Jason Cantarella}
\email{jason.cantarella@gmail.com}
\affiliation{Department of Mathematics, University of Georgia, Athens, GA 30602} 
\affiliation{Isaac Newton Institute, University of Cambridge, Cambridge, CB3 0EH, UK}
\author{Thomas W. Kephart}
\email{tom.kephart@gmail.com}
\affiliation{Department of Physics and Astronomy, Vanderbilt University, Nashville, TN 37235} 
\affiliation{Isaac Newton Institute, University of Cambridge, Cambridge, CB3 0EH, UK}
\author{Eric Rawdon}
\email{ericrawdon@gmail.com}
\affiliation{Department of Mathematics, University of St.~Thomas, St.~Paul, MN 55105 } 
\affiliation{Isaac Newton Institute, University of Cambridge, Cambridge, CB3 0EH, UK}
\date{\today}
\begin{abstract}
We model the observed $J^{++}$ mesonic mass spectrum in terms of energies for tightly knotted and linked chromoelectric QCD flux tubes.
The data is fit with one and two parameter models.
We predict a possible new state at approximately $1190$ MeV and a plethora of new states above $1690$ MeV.
\end{abstract}
\preprint{NI12089-TOD} 
\maketitle

\section{Introduction}

In 1867 Lord Kelvin suggested~\cite{LK} that elementary particles can be knots.
While his idea of knotted fluid vortices in the aether as fundamental objects of nature was revolutionary for his time, our present experimental knowledge does not agree with this conjecture.
Specifically, we now know that knotted fluid vortices are unstable and worse still, the aether does not exist.
Nevertheless, the idea is attractive for its simplicity of relating fundamental physical and mathematical objects, and it should not be discarded out of hand before being tested on various other physical systems. 
One such system where stable knotted configurations may exist is quantum chromodynamics (QCD), the subject of the present study.

Observations from many experiments can be interpreted as signatures of unusual mesonic states, i.e., bosonic hadrons that are not pure $q\bar{q}$~\cite{PDG}.
Such states can be broadly divided into the following types: (1)\ hybrids---bound states of quarks and gluons, like $q\bar{q}G$ with quantum numbers $J^{PC}=0^{-+}$, $1^{-+}$, $1^{--}$, $2^{-+}$, $\ldots$; (2)\ exotics---for example, four and six quark states, such as $qq\bar{q}\bar{q}$ and $qqq\bar{q}\bar{q}\bar{q}$ with quantum numbers $J^{PC}=0^{--}$, $0^{+-}$, $1^{-+}$, $2^{+-},\ldots$; (3)\ glueballs---states with no valence quarks at all, composed of pointlike or collective glue, e.g., string loops  \`{a} la Nielsen-Olesen~\cite{Nielsen-Olesen}, or closed flux tubes.
(Even though glueballs do not contain valence quarks, there are certainly sea  (virtual) quarks within a glueball.)
Glueballs are among the most studied and least understood classes of particles in QCD~\cite{Swanson}.
Lattice calculations, QCD sum rules, electric flux tube models, and constituent glue models lead to a consensus that the lightest valence quark-free state is a glueball with quantum numbers $J^{++}=0^{++} $~\cite{West}.
On the lattice, usually only  a single glueball state below $\sim$2 GeV is considered, since all the excitations are expected to be above this energy~\cite{Morningstar:1999rf}, however, a full study of topological operators responsible for knots and links on the lattice is computationally challenging at present and has not yet been carried out.
Nevertheless, studying such configurations would be interesting and potentially important for a better understanding of QCD~\cite{comment6}.

Although glueballs, and the $f$ states they are associated with, are one of the most widely discussed problems in hadronic physics, and, while many glueball models have been proposed~\cite{glueball refs}, there is still no consensus  of what constitutes a glueball beyond its $J^{++}$ quantum numbers. Here we take an  egalitarian approach. Specifically, we model all $J^{++}$ mesonic states, i.e., all $f_{J}$ and $f'_J$ states listed by the Particle Data Group (PDG)~\cite{PDG} ($f$ states, for brevity), as knotted or linked chromoelectric QCD flux tubes~\cite{comment1}. Hence we will use the term ``glueball'' loosely as a shorthand for any $J^{++}$  state in QCD.

\section{Model  }

In~\cite{Buniy:2002yx} two of us argued how to generalize various classical ideas from plasma physics to a semiclassical model of knotted and linked configurations in QCD.
To this end, we first recall that Maxwell's equations for an ideal plasma imply that flux lines are locked into the plasma flow.
This means that if the flux lines are knotted or linked, then the flux line topology is conserved as the flow evolves.  A classical consequence  of conservation of topology is
 the concept of helicity and its conservation in a plasma~\cite{Woltier}.
Helicity in this context corresponds to the degree of Gaussian linking of QCD flux tubes.
Keeping only the helicity fixed, while allowing other topological changes in flux lines, and minimizing the energy leads to the so-called Taylor states \cite{Taylor} in plasma physics. Keeping flux tube topology fixed leads to tight knots \cite{tight} in QCD.

The simplest configuration of linked flux tubes has the form of a Hopf link (denoted $2_1^2$) in which two unknots are linked together in the simplest way such that the Gaussian linking number  is one.
This is the tight Hopf link, in which fixed diameter $d$ flux tubes, carrying one  flux quantum each,  have the shortest total length.
The ratio of the total length of the tubes to their diameter $d$ is invariant for all such tight Hopf links.
We define the ``knot energy''  of the tight configuration $K$ as its dimensionless length   $l_K$, 
$$\epsilon_0(K)=l_K/d,$$
where in this example $K =2_1^2$, so then $\epsilon_0(2_1^2)=2(2\pi d)/d=4\pi$.  By ``knot energy,'' mathematicians generally mean some kind of repulsive electrostatic energy such as O'Hara's energy \cite{OH} or the Freedman-He-Wang \cite{FHW}  ``Mobius energy.''  Our  
$\epsilon_0(K)$ is called the ropelength of the configuration and the configurations which minimize the ropelength are called ``tight'' or ``ideal'' knots \cite{knotenergy}; see also \cite{RR}. For the Hopf link and a family of other simple links, where the tight configuration is known, see \cite{CKS}.
As expected, for the Hopf link this configuration consists of two linked circles passing through each other's centers.

The Hopf link is the simplest link, but there is an infinite family of topologically different link types. Many of these configurations have been tabulated by mathematicians \cite{DH,HT} 
and there are different topological invariants (such as linking numbers) which distinguish them. However, regardless of which invariants are used to identify the configurations, the ``frozen-in field'' hypothesis implies that tube topology is conserved. Hence, we conjecture that the ground states of all systems of flux tubes, with any type of nontrivial linking, are the states with the shortest length tubes.

Similar to linked flux tubes, we also consider self-linked (knotted) flux tubes, where 
the associated tight knot states and dimensionless knot lengths are  analogously defined.
The simplest example of a nontrivially knotted flux tube has the form of the trefoil knot $3_1$. 
This configuration has knot energy which has been numerically calculated to be  $\ve_0(3_1)\approx 16.3715$ \cite{PP}. Note that there are no known analytic forms for the lengths of any tight knots or links with nonplanar elements.

To begin the description of the model, we consider a high energy hadron-hadron collision in the process of rehadronization, where there are baryons, mesons and quantized fluxes confined to tubes. If the tubes are open, with quarks and anti-quarks at their ends, then they are excited baryon or meson states. Our interest is in closed tubes which can be self-linked (knotted) or linked with each other.
As a key part of our model, we  identify all the $f$ states as knotted or linked QCD chromoelectric flux tubes.
The topological quantum numbers (or knot/link type of the configuration) are what stabilizes the knotted and linked configurations, so we assume that non-topological (i.e., unknotted/unlinked) $J^{++}$ closed flux tube configurations are too unstable to have measurable widths.
A configuration with tightly knotted or linked flux tubes in the form of the knot or link $K$ will be called $f(K)$.
Note that topological invariants in QCD typically require instanton and Chern-Simons terms for their full description \cite{FF+FA}, but we will not need these subtleties here.

As argued above, nontrivial knotting and linking leads to  quasi-stable generalized minimum energy states. This implies the following theorem. 

\begin{theorem}
For a configuration with a topological charge measured by knotting or linking of flux tubes of a constant radius (due to quantized flux, and therefore of constant energy or mass per unit length), the generalized minimum energy state is the one that minimizes $\epsilon(K)$ and therefore ties the knot or link with the shortest tube. 
  \label{theorem}
\end{theorem}

Clearly, the minimum energy corresponds to the minimum dimensionless length of tube needed to support the topology.
Since the length and energy coincide up to a rescaling, the proof proceeds trivially by inspection. 
(Note that  this proves a quantum analog of Moffatt's 1985 conjecture that higher order linking leads to
positive lower bounds on configuration energy.)
We will see below that the approximation of a fixed energy per unit length can be improved by an analysis 
of the effect of field  rearrangement within a bent tube.

We conclude that the quantum case of tight flux tube configurations is much simpler than the corresponding classical case where one minimizes energy with a flux constraint.
However, the quantum case suggests that it may be possible to sum or integrate any large number of flux quanta to get the classical result for the generalized minimum energy of a Gaussian-linked or higher order topological configuration. (Such a result would complete the proof of Moffatt's conjecture for the classical case.) This is a side issue from our main purpose here that will be explored elsewhere.

Let us now proceed with the further description of our phenomenological model, which is similar to the model in \cite{Buniy:2002yx}, but with some minor modifications.
Pulling a quark-antiquark ($q{\overline q}$) pair apart in the QCD vacuum, a chromoelectric flux tube forms along a path connecting the quark and antiquark.
If the $q$ is annihilated against the ${\overline q}$, or if it is annihilated by a ${\overline q}$ in another $q{\overline q}$ pair, where the new $q$ is in turn annihilated, etc., to close the path, then a closed flux tube containing one flux quantum can form as an unknot or a knot.
   A flux tube following such a curved path could also, for instance, be due to multiple scattering of the inital $q{\overline q}$ pair before its mutual annihilation.
If the closed tube is a knot, or if it ends up being linked with another closed tube, then such an object has at least one nonzero topological quantum number, and this quantity is what tends to stabilize the tight configuration, which we either identify with one of the observed $f$ states or use to predict a new state.
To allow hadronization to run its course, we assume that the typical time scale needed to reach a tight configuration is shorter than the typical lifetime of the hadron.

Since the publication of \cite{Buniy:2002yx} in 2003, there has been slow continuous physics progress in the refinement of the  $f$ states data as summarized by the PDG~\cite{PDG}, with smaller error bars from better statistics and a few new states now listed in their summary tables.
The change in the $f$ state data that affects our model the most is the PDG's realignment of the mass of the $\sigma$, formerly called the $f_0(600)$ and now reassigned as the $f_0(500)$.
While there has been no new data since 2007, there has been several new reanalyses  of compilations of the existing data.
The 2010 PDG value of the $f_0(600)$ mass was  reported to be in the range $400-1200$ MeV and we previously used $800 \pm 400$ MeV as our approximation.
Now the PDG is reporting a mass range for the $f_0(500)$ of $400-550$ MeV.
This is a drastic change since the range has contracted by a factor of 5 and the central value has dropped by over 300 MeV.

Contrary to the rather gradual physics progress, the mathematical knowledge of tight knots has changed dramatically from what was used in \cite{Buniy:2002yx} to what it is today~\cite{ridgerunner}.
In 2003 the lengths of only a handful of tight knots and non-planar links were known, and some of those only to an accuracy between 5\% and 10\%.
Now we know the complete length spectrum of the first several hundred tight prime knots and non-planar links with an accuracy that is assumed to be in the $0.1\%-1.0\%$ range for most of this spectrum. The lengths of the composite knots have recently appeared~\cite{CLR}. We present new computations covering the final piece of the puzzle---ropelengths of composite \emph{links}---for the first time below. The current limitation on knot energies needed for the model is due to the fact that we are dealing with physical knots and links that can be constricted or distorted (see below), thus increasing the errors on the effective lengths.
Even with this caveat, we still can advantageously refit the $J^{++}$ data by comparing it with high accuracy tight knot and link data after adding estimated errors, all of which is collected in Tables \ref{table-lengths1} through \ref{table-cclengths2} below.

Let us summarize the model assumptions:
\begin{enumerate}
  \item There is a one-to-one correspondence between $f$ states and tightly knotted and linked chromoelectric flux tubes.
  \item The flux is quantized with one flux quantum per tube.
  \item Knotted and linked flux tubes are stabilized by topological quantum numbers.
  \item The tube diameter is in the $\sim 0.1$ fm range. (This corresponds to a string tension of approximately $ 400$ MeV, which agrees with lattice estimates.)
  \item The quantity $J$ in an $f_J$ or $f'_J$ state is the intrinsic angular momentum of the associated knotted solitonic solution of the QCD field equations.
  \item The relaxation to a tight state configuration (via processes where no topology change is involved) is faster than its decay rate (via processes with topology change) for an $f$ state, i.e., $\tau_{\textrm{relax}}\ll \tau_{\textrm{decay}}$.
\end{enumerate}

One modification from \cite{Buniy:2002yx} is that we now assume $J$ is the intrinsic angular momentum rather than the rotational angular momentum. 
We do this because the tube diameter is now assumed to be smaller and hence the rotational energy level spacing to be larger, $\sim 500$ MeV, as opposed to a few MeV for the thicker tubes assumed in \cite{Buniy:2002yx}. The other significant modification is that we correct the energy due to tube curvature as discussed in the next section and include estimated errors due to other physical corrections.

\begin{figure}
\includegraphics[width=400pt]{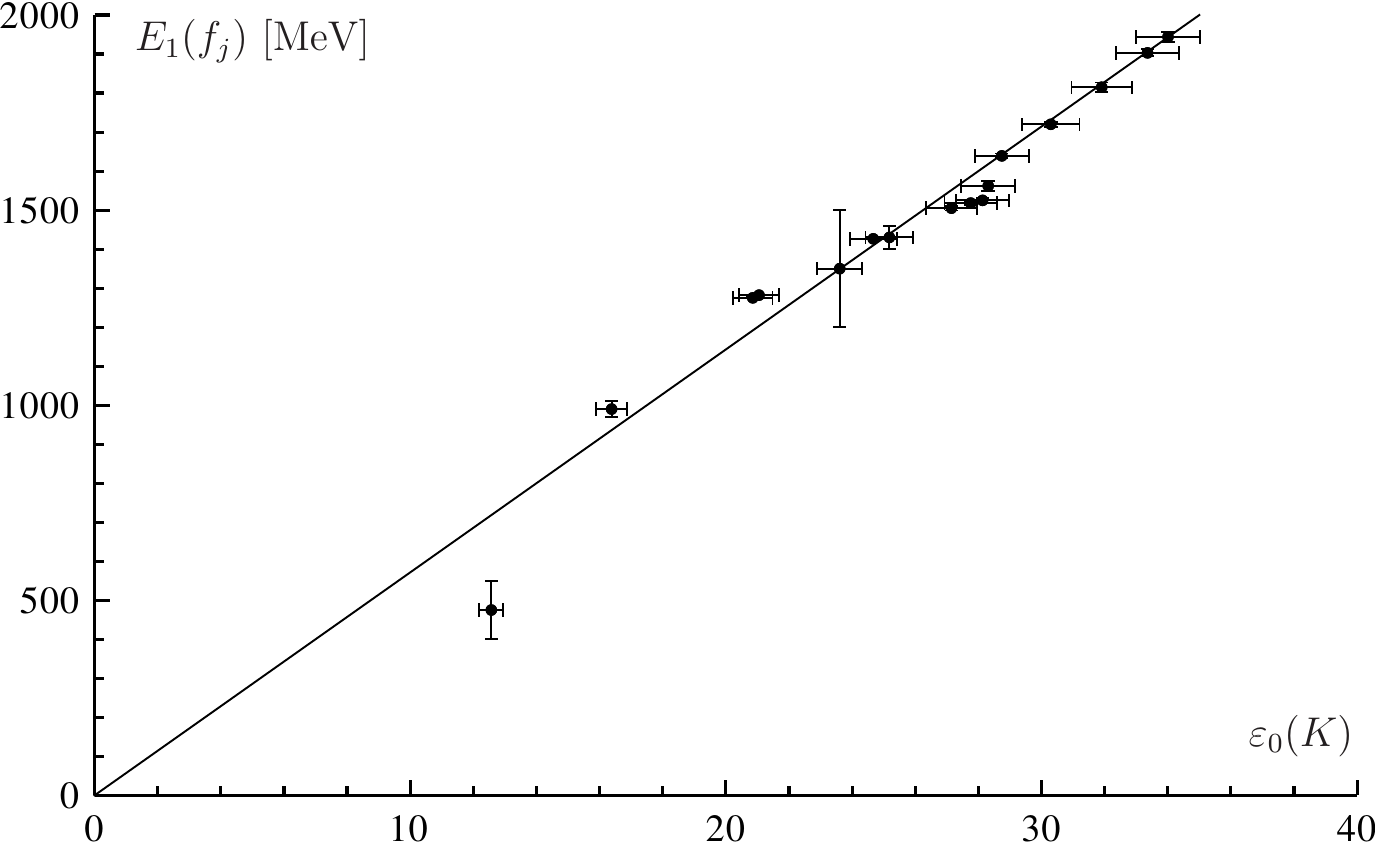}
\caption{\label{1p-fit} Fit using uncorrected knot/link lengths:
Fit of the $f_J$ states data to tight knot and link lengths (ropelengths).
Errors are shown for the states, but they are too small to be visible for the lengths of ideal knots and links, however we include a 3\% error in the knot energies due to the fact that we are dealing with physical knots and links. (See  discussion in the text.) Non-fitted knots and links are not shown.}
\end{figure}

We begin by identifying ideal knot and link lengths    with glueballs and/or predicted glueballs, where we include all $f$ states.
The lightest candidate is the $f_0(500)$, which we identify with the shortest knot or link, i.e., the Hopf link $2^2_1$; the $f_0(980)$ is identified with the next shortest knot or link, in this case the trefoil knot $3_1$, etc. 

Our initial one-parameter fit of the data is shown in Figure \ref{1p-fit}.
The slope is $\Lambda_\textrm{tube}=57$ MeV  and  $\chi^2=84$. The fit is poor mainly because of the constraint imposed by the very small error bars on the masses of the $f_2(1270)$ and the $f_1(1285)$. We will now see how to improve the fit when ideal tubes are replaced with physical tubes. Note we are already assuming errors of 3\% on the knot lengths which anticipates this replacement.
 
\section{Curvature corrections}

In the discussion so far we have assumed uniform flux across the cross section of the tubes, but flux is not necessarily uniform over the cross section of curved tubes. This leads us to define a new energy functional for tubes which we call the ``flux tube energy.''

\begin{figure}
\includegraphics[width=200pt]{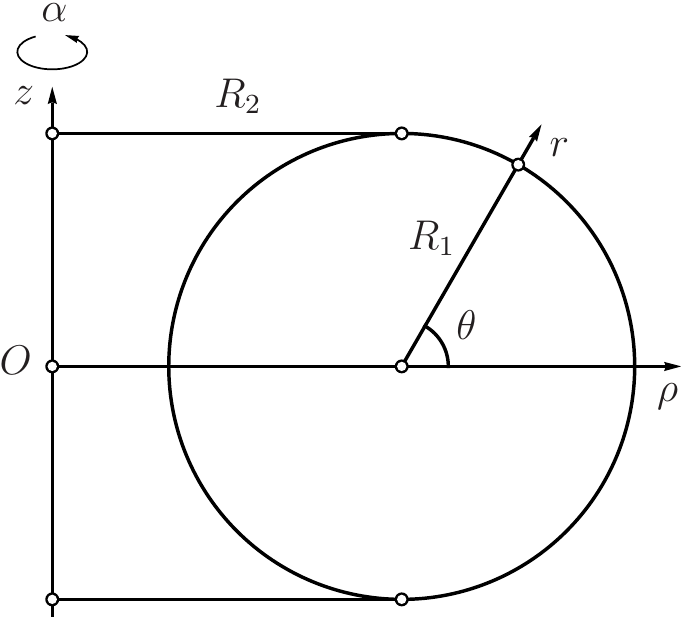}
\caption{\label{coordinates} Cross section of a toroidal flux tube of minor radius $R_1$ and major radius $R_2$. Both polar and cylindrical coordinates are shown.}
\end{figure}

To motivate our definition, we consider the effect of bending on the total energy of a field confined to a tube. First recall that the magnetic field of an ideal toroidal solenoid with fixed flux falls like $1/\rho$ from the symmetry axis. To see this, choose cylindrical coordinates $(z,\rho,\alpha)$ as shown in Fig. \ref{coordinates} and note that symmetry requires the field be in the $\alpha$ direction.
(For an elementary argument see \cite{GriffithsEM}.)
Here we will proceed via a variational argument which is a simpler alternative.

We hold the flux $\Phi$ fixed and vary the field  to find the  
functional form of the energy $W$ for a toroidal solenoid. The general form of the energy is
\begin{align}
W = \frac{1}{2}\int_D B^2  \rho \,dz\,d\rho\, d\alpha.
\end{align}
The $dzd\rho$ integration runs over the cross section of the tube $D$. 
Since $B$ is independent of $\alpha$, the $\alpha$ integration gives
\begin{align}
  W={\pi}\int_D B^2 \rho\, dz\, d\rho.
  \label{}
\end{align}
The flux through $D$ is 
\begin{align}
\Phi=\int_D  B\,dz\,d\rho.
\end{align}
Now we want to vary $W$ with respect to $B$ while holding $\Phi$ fixed.
This is equivalent to considering
\begin{align}
  \delta(W-\lambda\Phi)=0,
\end{align}
where $\lambda$ is a Lagrange multiplier.
For unit vector $n$ normal to the cross section, the variation of $B$ gives
\begin{align}
  \int_D(2\pi B\rho-\lambda n)\cdot \delta B\,dz\,d\rho=0,
\end{align}
which  vanishes for arbitrary $\delta B$ only if
\begin{align}
B(\rho)=\frac{\lambda n}{2\pi \rho}.
\end{align}
We find $\lambda$ from the requirement $\Phi=\textrm{const}$, which gives 
\begin{align}
  &B=\frac{\Phi}{\rho I},\\
  &W=\frac{\pi\Phi^2}{I}
\end{align}
  with
  \begin{align}
  &I=\int_D\frac{dz\,d\rho}{\rho} .
\end{align}

\begin{figure}
\includegraphics[width=300pt]{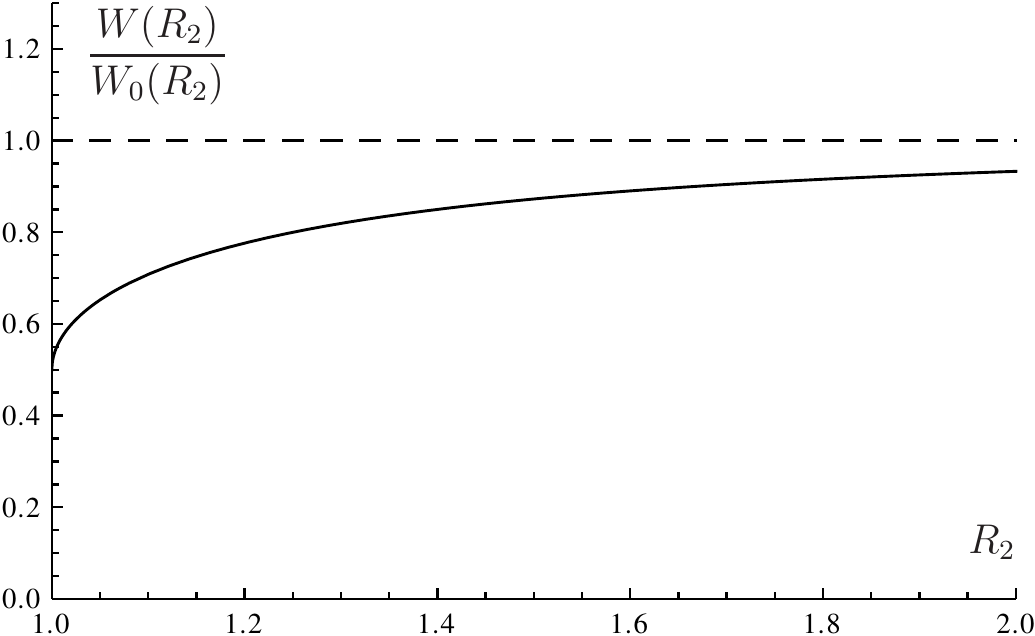}
\caption{\label{w} The function $W(R_2)/W_0(R_2)$ (solid curve) and its approximation (dashed line) for $R_2/R_1\gg 1$ and $R_1=1$.}
\end{figure}

To calculate the integral $I$ over the cross section of a torus of major radius $R_2$ and minor radius $R_1$, it is convenient to introduce polar coordinates $(r,\theta)$ with the origin at the center of disk $D$, plus a toroidal angle $\alpha$; see Fig. \ref{coordinates}.
The result of integration over $D$ is
\begin{align}
  &I=2\pi\left[R_2-\sqrt{R_2^2-R_1^2}\right],
  \label{}
\end{align}
which leads to
\begin{align}
  W(R_2)=\frac{\Phi^2 }{2\left[R_2-\sqrt{R_2^2-R_1^2}\right]}.
\end{align}
The analogous result for the cylinder of length $2\pi R_2$ is 
\begin{align}
  W_0(R_2)=\frac{\Phi^2R_2}{R_1^2},
  \label{}
\end{align}
and so the ratio
\begin{align}
  \frac{W(R_2)}{W_0(R_2)}=\frac{(R_1/R_2)^2}{2[1-\sqrt{1-(R_1/R_2)^2}]}, 
  \label{}
\end{align}
the graph of which is plotted in Fig. \ref{w}, will define our new energy. 

Formally, for an embedded tube $K$ of fixed radius $R_1$ and parametric centerline curve $\gamma(s)$ with curvature $\kappa(s)$, we define the flux tube energy $\ve(K)$ by the  integral
\begin{align}
\ve(K) = \frac{1}{2\pi R_1^2} \left(L +\int_0^L   \sqrt{1 - R_1^2 \kappa^2(s)}\,ds \right) , 
\label{TubeEnergy}
\end{align} 
where $L=\int_\gamma   ds$ is the length of the center line.  (In numerical studies of tight knots and links, it is observed that the integral in (\ref{TubeEnergy}) is typically $\sim \frac{\sqrt{3}L}{2}$ which translates into an $\sim$7\% correction of the energy from the ropelength value.)
It would be an interesting project to numerically minimize the flux tube energy for various knot and link types. However, this is likely to be a somewhat challenging project: minimizing functionals of curvature (a second derivative of position) constrained by tube contact (a function of position) is quite difficult. Still, there has been recent progress in the numerical modeling of elastic rods with self-contact \cite{elasticrods} which lead to us hope that these computations may be tractable in the near future. 

\begin{figure}
\includegraphics[width=300pt]{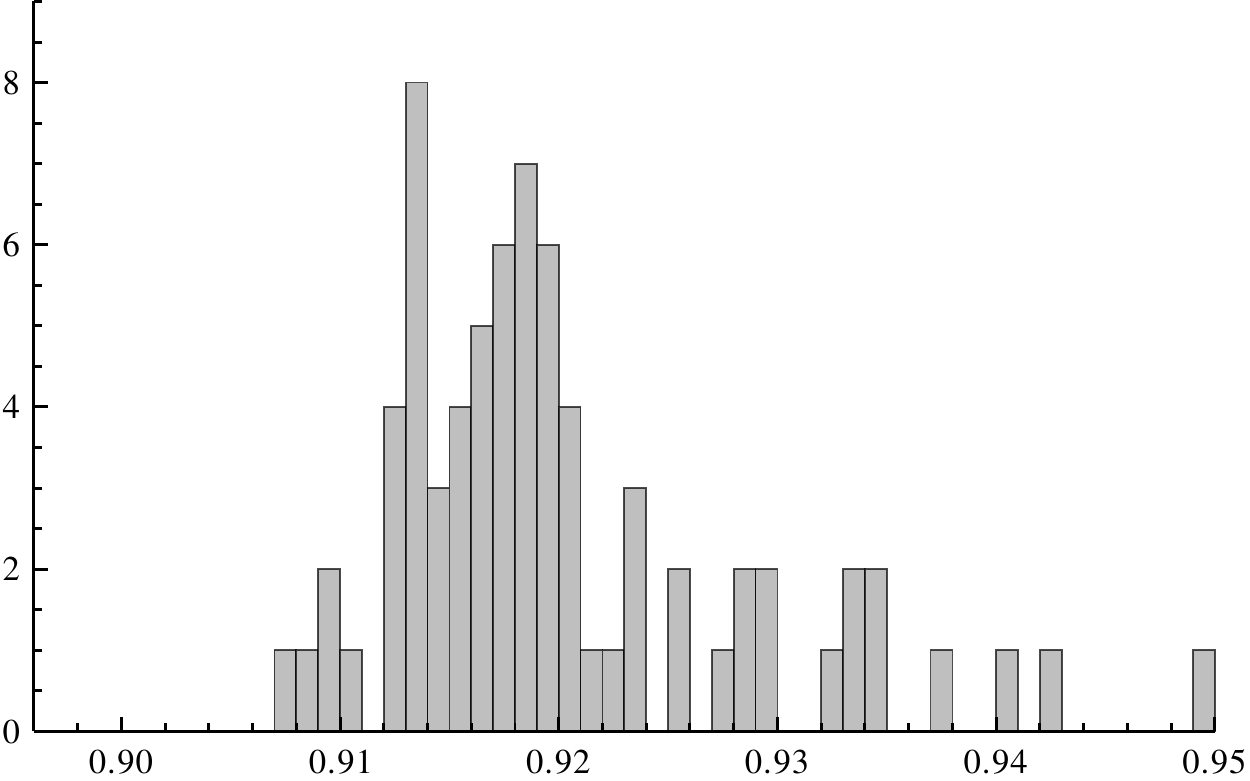}
\caption{\label{smallhistogram}Histogram of the magnitudes of the curvature corrections to the first 72 knot and link lengths.}
\end{figure}

\begin{figure}
\includegraphics[width=300pt]{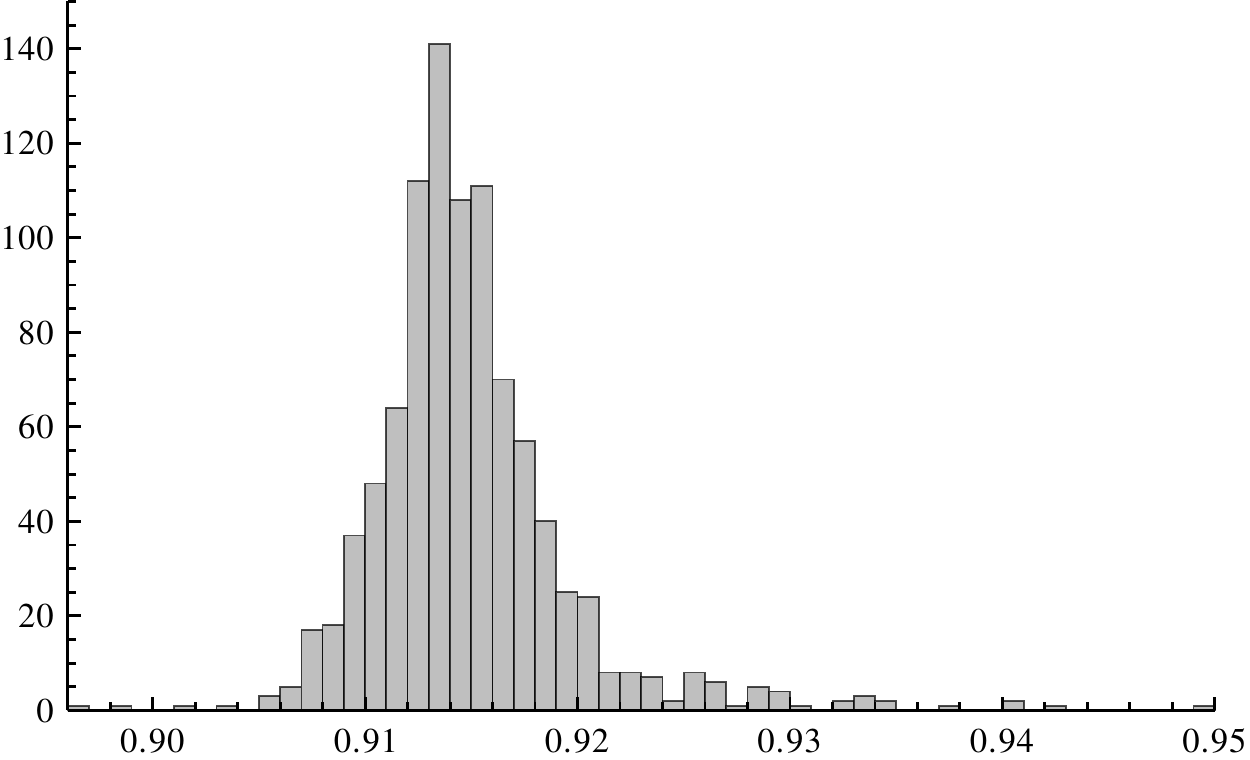}
\caption{\label{largehistogram}Histogram of the magnitudes of the curvature corrections to all $945$ currently tabulated knot and link lengths.}
\end{figure}

In the meantime, we have chosen to minimize the original $\epsilon_0(K)$ energy numerically using  \texttt{ridgerunner}  \cite{ridgerunner}, and then compute the $\ve(K)$ energy for the $\epsilon_0(K)$-minimizing configurations on the grounds that the difference between $\ve(K)$-minimizing and $\epsilon_0(K)$-minimizing configurations are likely to be small. In Figure \ref{smallhistogram} we have histogrammed the shortest 72 knots and links after curvature corrections have been applied.   In Figure \ref{largehistogram} we have histogrammed the currently available complete set of 945 curvature corrected knots and links.

We should remark that in some cases the curvature is discontinuous, but the field is not. For example, the inner loop of the chain of three elements (the $2^1_1\#2^1_1$) is shaped like a race track---two straight sections  connected by two half circles. In a solenoid of this form the curvature is discontinuous at the junctions, but if we move along the field lines we find that the fields are already changing before they reach the junctions since the windings are different in the  regions beyond the junction and the field is affected by the fringe fields in that region. Hence the fields can be continuous through the junction.


As an example of a case with both length and curvature correction that can be calculated exactly \cite{CKS}, consider the chain of three unknots $2^2_1\# 2^2_1$ 
which has length $6\pi+2\approx 20.8496$ and has $R_2=2R_1$ in curved regions and $R_2\rightarrow \infty$ in straight sections.
We find an overall corrected value 
\begin{align}
  \ve(2^2_1\# 2^2_1)=\frac{1}{4(2-\sqrt{3})}6\pi+2\approx 0.933013(6\pi+2)\approx 19.4529
  \label{}
\end{align}

Note that all exactly calculable link lengths in our tables are unique, but degeneracies can occur at longer lengths. The first such examples are the links corresponding to the $E_6$ and $D_6$ Dynkin diagrams. Both have the length $\epsilon_0(E_6)=\epsilon_0(D_6)=12\pi +7$ (which falls slightly beyond the largest lengths included in our tables) and both have the same curvature corrected energy $\ve(E_6)=\ve(D_6)=\frac{3\pi}{(2-\sqrt{3})} +7$. Other corrections should lift this degeneracy. 

 \section{Results}

In our model, the chromoelectric fields~\cite{comment3} $F_{0i}$ are confined to the knotted and linked tubes, each carrying one quantum of conserved flux~\cite{flux,soliton}.
We consider a stationary Lagrangian density
\begin{align}
  {\cL}=\frac{1}{2}\tr F_{0i}F^{0i}-V,
  \label{}
\end{align}
where, similar to the MIT bag model~\cite{MIT-bag}, we included the possibility of a constant energy density $V$.
To account for conservation of the flux $\Phi_E$, we add to $\cL$ the term
\begin{align}
  \tr\lambda[\Phi_E/(\pi a^2)-n^iF_{0i}],  
  \label{}
\end{align}
where $n^i$ is the normal vector to a section of the tube of radius $a$ and $\lambda$ is a Lagrange multiplier.
Varying the full Lagrangian with respect to $A_\mu$, we find
\begin{align}
  &D^0(F_{0i}-\lambda n_i)=0,\\
  &D^i(F_{0i}-\lambda n_i)=0,
  \label{}
\end{align}
which have the constant field solution
\begin{align}
  F_{0i}=(\Phi_E/\pi a^2)n_i.
  \label{}
\end{align}

With this solution, the energy is positive and, to first approximation, proportional to the length of the tube $l$ and thus the minimum of the energy is achieved by shortening $l$ (i.e., tightening the knot), subject to the curvature correction discussed above and other corrections discussed below.

We proceed to identify knotted and linked QCD flux tubes, i.e., curvature corrected physical flux tubes, with glueballs and/or predicted glueballs, where we include all $f$ states.
The lightest candidate is the $f_0(500)$, which we identify with the shortest curvature corrected knot or link, i.e., the Hopf link $2^2_1$; the $f_0(980)$ is identified with the next shortest knot or link, in this case the trefoil knot $3_1$, etc. By the fourth knot/link, the ordering begins to be reshuffled due to the curvature corrections, see Tables \ref{table-lengths1} and \ref{table-cclengths1}.

All knot and link lengths have been calculated for states corresponding to energies well beyond $2\,\textrm{GeV}$.
Above $\sim 2\,\textrm{GeV}$ the number of knots and links grows rapidly, and so the corresponding hadronic states should become dense relative to their typical width.
Hence we will confine our investigations to knot lengths corresponding to all known $f$ states below $\sim 2$ GeV.

\begin{figure}
\includegraphics[width=400pt]{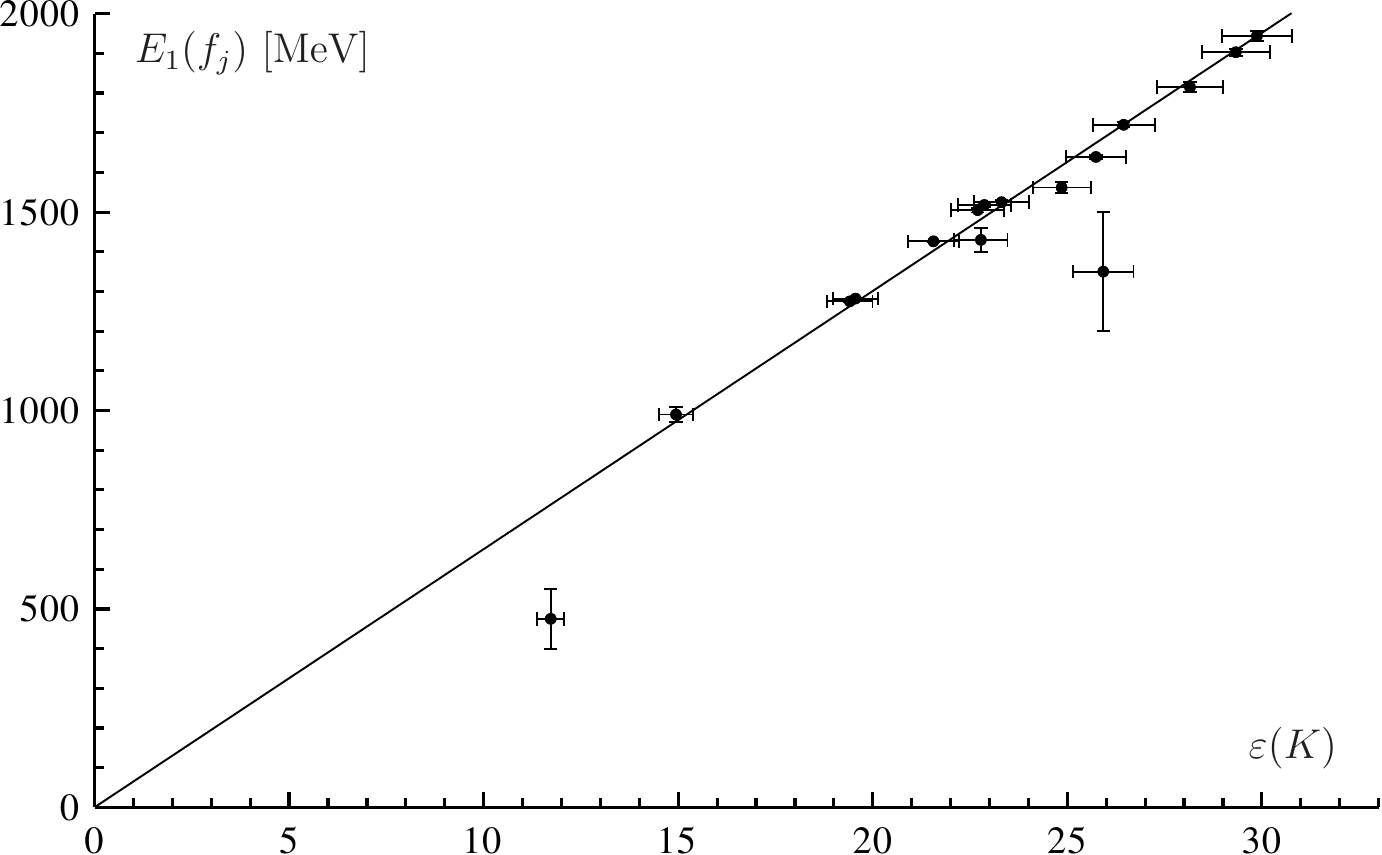}
\caption{\label{cc-high-1p-fit} Fit using curvature-corrected lengths:
High one parameter $f_0(1370)$ fit of the $f_J$ states data to the curvature-corrected knot and link data.
Errors are shown for the states, and the 3\% estimated from the text is shown for the knots and links. Non-fitted knots and links are not shown in  this figure.}
\end{figure}

\begin{figure}
\includegraphics[width=400pt]{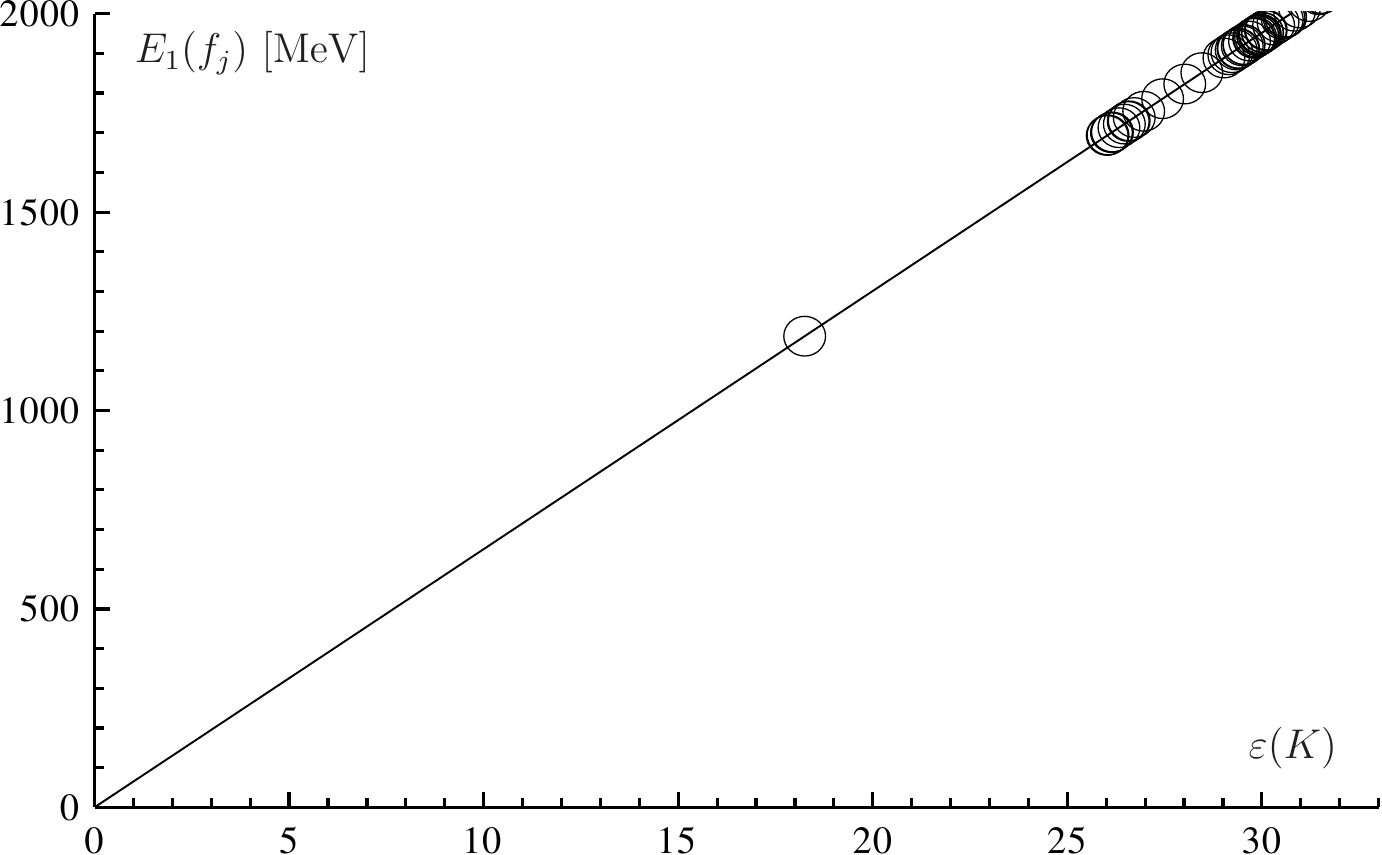}
\caption{\label{cc-high-1p-prediction} Predictions using curvature-corrected lengths:
The circles are locations of knots and links that do not have corresponding $f_J$ states in Fig.~\ref{cc-high-1p-fit}.
Hence these are the locations of states predicted by the high $f_0(1370)$ one parameter fit of the model.
}
\end{figure}
\begin{figure}
\includegraphics[width=400pt]{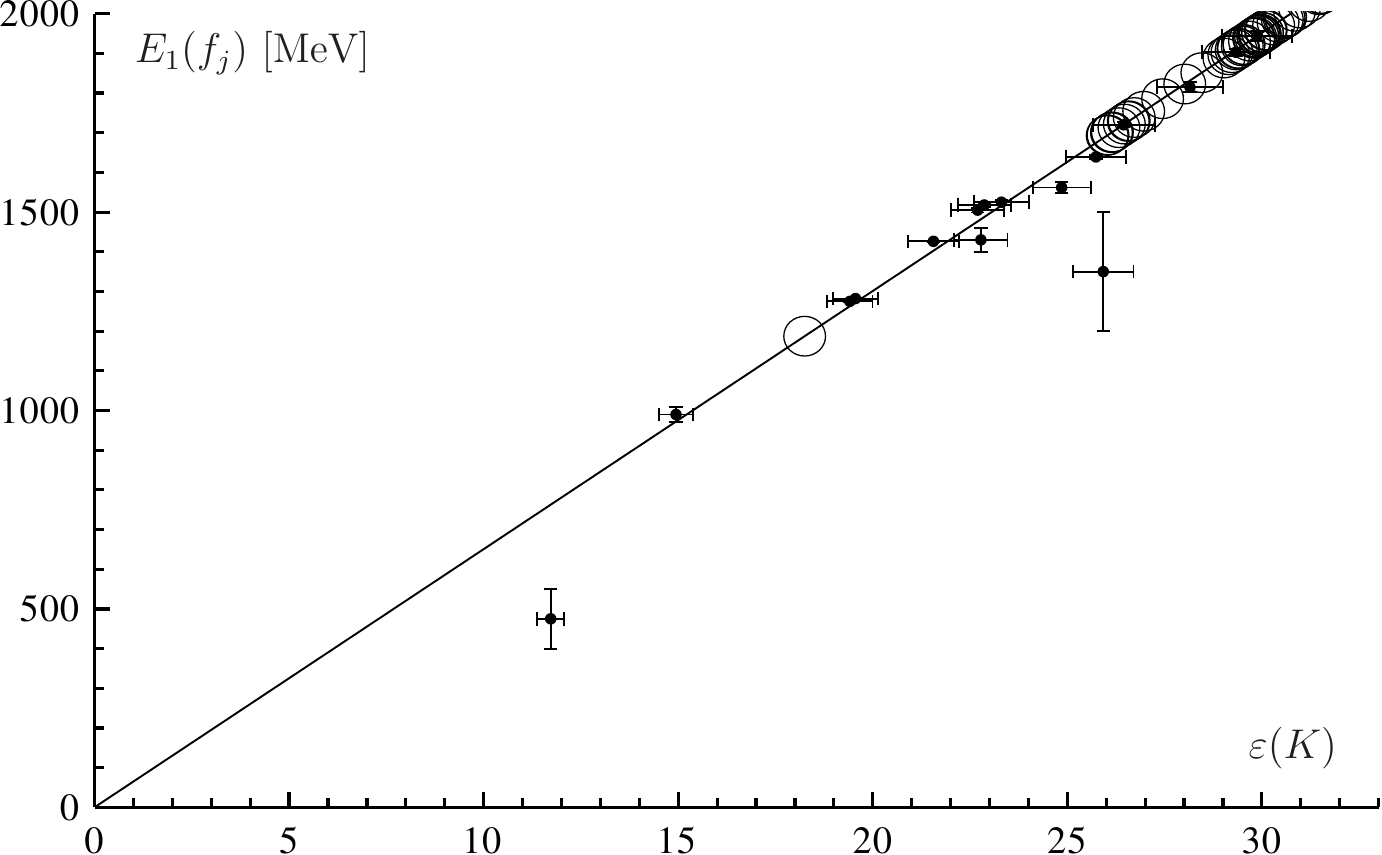}
\caption{\label{cc-high-1p-fit-prediction} Curvature-corrected lengths:
The combined set of measured and fitted states (dots with error bars) and predicted states (circles) for the high $f_0(1370)$ one parameter fit.}
\end{figure}

\begin{figure}
\includegraphics[width=400pt]{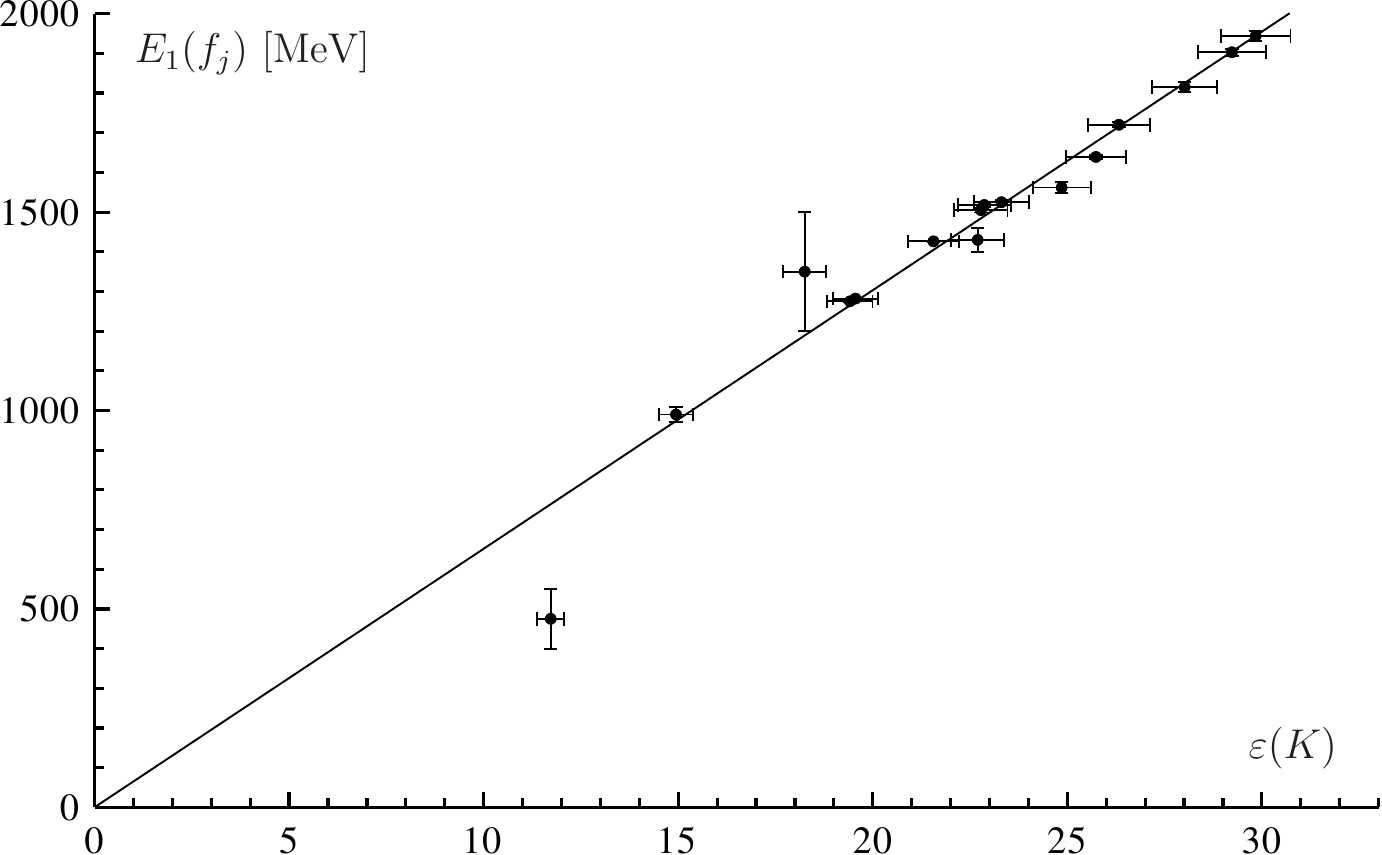}
\caption{\label{cc-low-1p-fit} Low $f_0(1370)$ one parameter fit with curvature-corrected lengths:
This is our best fit of the $f_J$ states data to the knot and link data.
Errors are shown for the states, as is the 3\% error estimate of knot and link lengths. Non-fitted knots and links are not shown.}
\end{figure}

\begin{figure}
\includegraphics[width=400pt]{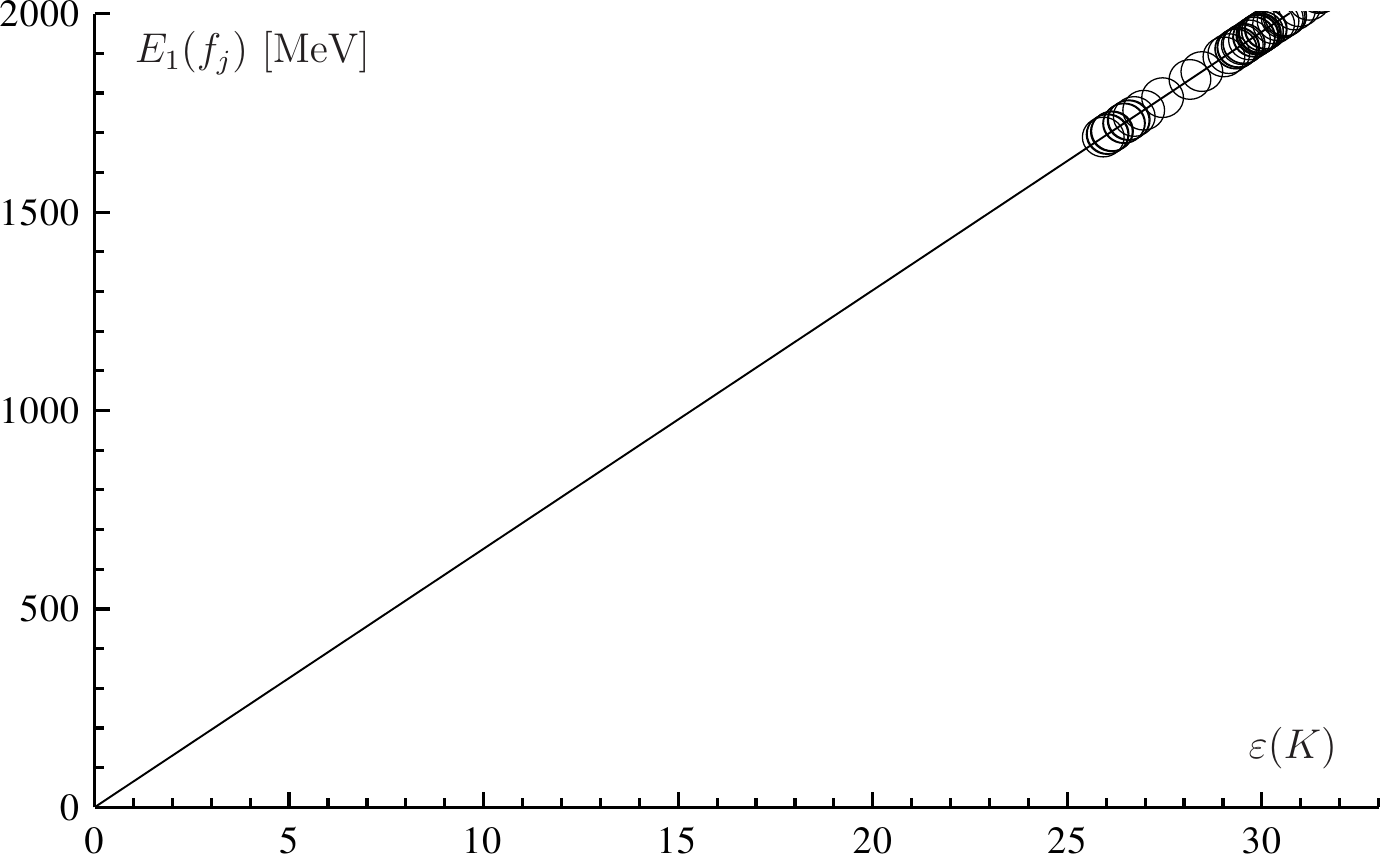}
\caption{\label{cc-low-1p-prediction} Low $f_0(1370)$ one parameter fit for curvature-corrected lengths:
This figure shows the knots and links that do not have corresponding $f_J$ states in Fig.~\ref{cc-low-1p-fit}.
Hence these are the locations of states predicted by the model.
}
\end{figure}

\begin{figure}
\includegraphics[width=400pt]{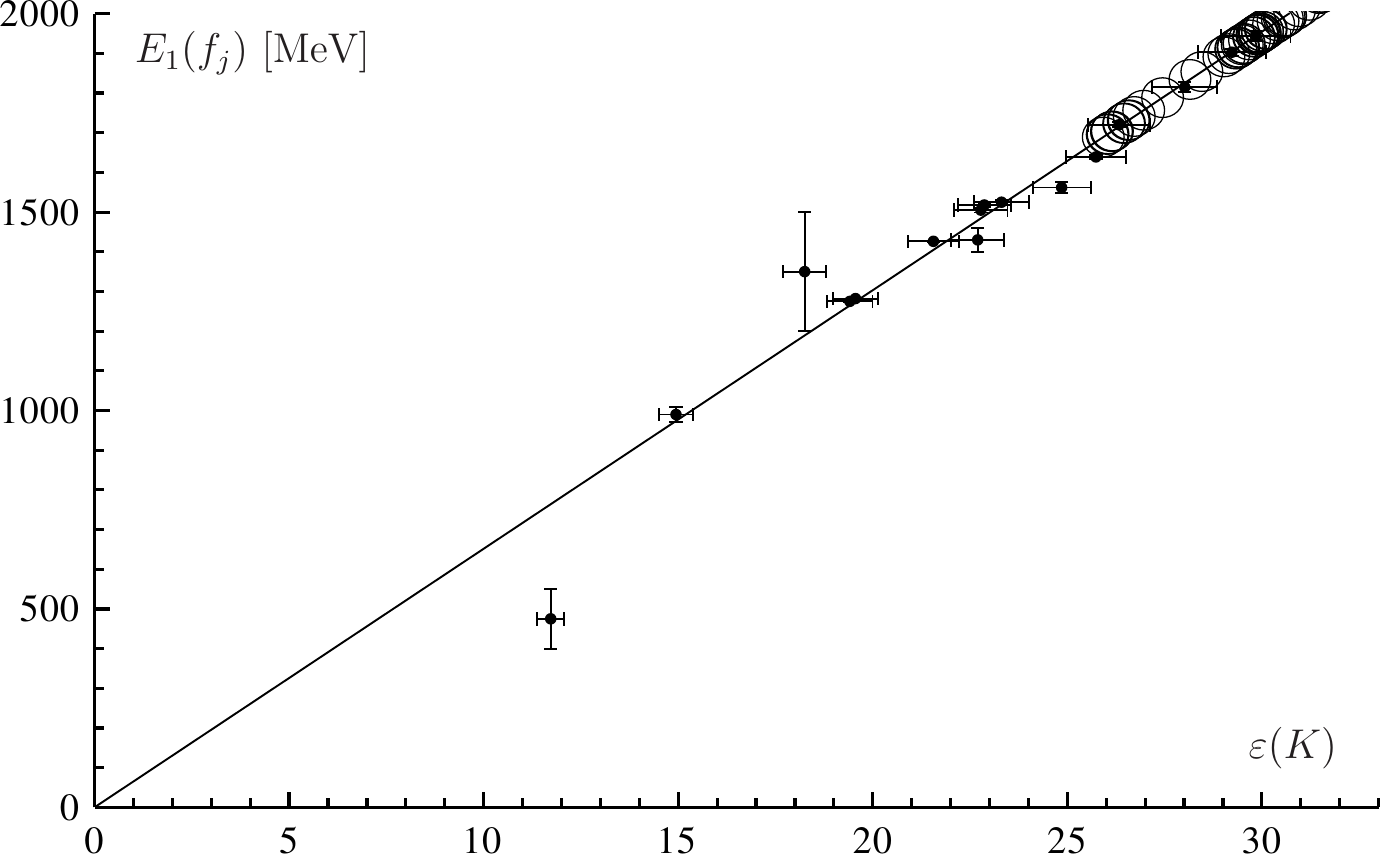}
\caption{\label{cc-low-1p-fit-prediction} Low $f_0(1370)$ one parameter fit for curvature-corrected lengths:
TThis figure shows the combined set of measured and fitted states (dots with error bars) and predicted states (circles).}
\end{figure}

Our detailed results are collected in Table \ref{table-lengths1} through \ref{table-cclengths2}, where we list the masses and error bars for the $f$ states (other properties can be found in \cite{PDG}) and our identifications of these states with knots and links together with the corresponding knot and link lengths (see Tables \ref{table-lengths1} and \ref{table-lengths2}), curvature corrected lengths (see Tables    \ref{table-cclengths1}, \ref{table-cclengths2} and \ref{table-cclengths1H}) and fitted energies.

We will give two interpretations of the data. The first possibility is with the $f_0(1370)$ identified with the $5_1$ knot which results in a prediction of a new state near 1190 MeV  identified with the $4^2_1$ link.
The other possibility, which gives our best fit, is to identify the $f(1370)$ itself with the $4^2_1$ link to give a one-to-one matching of the first 12 $f_J$ states with the first 12 knots and links. We will discuss the details of these options below.

For comparison purposes we have displayed results for knot energies proportional to lengths (Tables \ref{table-lengths1} and \ref{table-lengths2}) and also for the curvature corrected knot energies,  (see Tables \ref{table-cclengths1}, \ref{table-cclengths2} and \ref{table-cclengths1H}). Let us begin with the uncorrected length case.
In Figure \ref{1p-fit} we display a one-parameter least-squares fit to the experimental data (below $1945\,\textrm{MeV}$) for the mass spectrum of $f$ states identified with  knots and links. The fit  is
\begin{align}
  E(K)=\Lambda_{\textrm{tube}}\,\ve_0(K), \quad \Lambda_{\textrm{tube}}=57.14\pm 0.53\,\textrm{MeV},
  \label{fit1}
\end{align}
where $\ve_0(K)$ is the dimensionless length of the knot or link $K$ as defined above.
The fit \eqref{fit1} shows fair agreement with our model.
One measure of the quality of the fit is given by the adjusted $R$ squared, $R^2=.998$.
Since we have more knots than $f$ states, the relevance of $R^2$ must be carefully interpreted and not taken at face value, especially beyond $\sim1700$ MeV where the lack of particle data per knot becomes pronounced.

Compared to our 2003 results \cite{Buniy:2002yx}, the $f_0(500)$ now falls well below the line of our fit and we pay a penalty in $\chi^2$.
The result is also partially responsible for the change in the value of our fit parameter $\Lambda_{\textrm{tube}}$,  compared to what we find using the 2010 values for the $f_0(600)$. It will be interesting to see if the new $f_0(500)$ PDG numbers are stable. Since they are extracted from partial wave analysis and multiply subtracted dispersion relations, there are questions about comparing with the mass values for other $f$ states found in invariant mass plots. There is also the important issue of mixing with four quark and other resonances. However, we are not in a position to comment further on these matters with confidence except for the general qualitative remark that in a flux tube model we could expect mixing with excited mesons if there is resonant behavior of tube breaking (forming a $q\bar{q}$ pair) and rejoining (by annihilating the pair). Resonant breaking and rejoining a two positions, or in two link tubes mixes with four quark states, etc.

Next we consider the curvature-corrected case where the one parameter fit is shown in Figure \ref{cc-high-1p-fit}. The $\chi^2$ is substantially improved over the uncorrected length case and drops to $\chi^2=33$ to a large extent due to  the two most restricting states, the $f(1270)$ and the $f(1285)$, have been brought in line with the rest of the $f_J$ data by the curvature corrections. Hence we continue to consider only the curvature-corrected length case for the remainder of this section.

Figure \ref{cc-high-1p-prediction} shows the locations of knots and links with no corresponding $f$ state, hence it gives the locations of new states predicted by the model.
Our first new state is at $1190$  MeV, corresponding to the $4^2_1$ link.
It is interesting to consider the PDG entry for the state $f_2(1270)$.
Of the $36$ quoted experimental observations, all but one is within three $\sigma$ of the PDG average mass of 1275 MeV.
The $5.5\,\sigma$ outlier at $(1220\pm10)$ MeV is from the process $pp\rightarrow pp\pi^{+}\pi^{-}$ in the experiment of Breakstone et al.~\cite{Breakstone:1990at}, and we suggest its identification with our predicted state at $\sim1190$ MeV.

We further predict twelve states around $(1710\pm 20)$ MeV.
To justify such a proposal, one only need to look at the PDG entry for the state $f_0(1710)$ to see a considerable amount of tension in the data with a large number of incompatible mass measurements in this region. 
We interpret this as an indication of multiple $J^{++}$ states near $1700$ MeV that need to be resolved, just as is suggested by our model; see Tables \ref{table-lengths1} and \ref{table-cclengths1}.
Similar reasoning applies to the states in the vicinity of  $f_2(1810)$, $f_0(1910)$, and $f_2(1950)$ where there is also tension in the data.
A global statistical analysis of $f$-state data to establish a statistical significance of such an interpretation should be carried out.
Figure \ref{cc-high-1p-fit-prediction} gives both the combined fit and predicted masses and is displayed for convenience.
Better HEP data will provide further tests of the model and improve the high mass identification.

We are now ready to discuss the second fit possibility. Since the error on the $f_0(1370)$ mass is rather large ($\pm 150$ MeV) it can be identified with several different knots and links. However, the identifications of the other $f$-states, except for the $f_0(500)$, are much more constrained due to the small errors on their masses and this in turn restricts the identification of the $f_0(1370)$ to two allowed choices with reasonable $\chi^2$s. We call the case discussed above the high-fit, where we identify the $f_0(1370)$ with the $5_1$ knot, and the low-fit where we identify the $f_0(1370)$ with the $4^2_1$ link. The high-fit is the one that predicts a new state at $\sim 1190$ MeV, while the low-fit gives a one-to-one match between the first 12 $f$-states and the first 12 knots and links. The low-fit gives a somewhat better $\chi^2$, but the high-fit gives an acceptable $\chi^2$ and could be required if for instance the 1220 MeV state of \cite{Breakstone:1990at} is confirmed making it  necessary to free up a low energy knot/link to identify with it. Since we have already presented the high-fit above, we now proceed to discuss the low-fit.

The low-fit leaves no gaps in the spectrum until we get near 1700 MeV. The $\chi^2$ is improved and
the fitted value of $\Lambda_{\textrm{tube}}$ is similar to the high-fit case. We find
\begin{align}
  E(K)=\Lambda_{\textrm{tube}}\,\ve(K), \quad \Lambda_{\textrm{tube}}=65.16\pm 0.61\,\textrm{MeV}.
  \label{fit2}
\end{align}
The low-fit \eqref{fit2} is our best overall fit to the data. We have tabulated the fitted  energies for the first 72 knots and links, identifying all $f$-states of energy less than 2 GeV, hence giving predictions of many new states at 1690 MeV and above.

Now let us discuss the statistical significance of our results. First we have done a number of tests to determine if our hypothesis that glueballs are knotted flux tubes is likely to be correct. 
Let us begin with the well-known Kolmogorov-Smirnov test, which tests distribution functions ${\hat F}(y)$ and $F(y)$ by means of the quantity 
\begin{align*}
   \sup_y |{\hat F}(y)- F(y)|,
  \label{}
\end{align*}
which in this case are the distribution functions of the glueball masses and calculated knot and link lengths.
The resulting $p$-value for the Kolmogorov-Smirnov test is 
\begin{align}
  p_{\textrm{KS}}=0.95,
  \label{}
\end{align}
where we recall that $p$ is bounded $0\le p\le 1$ and $p<0.01$ implies poor correlation, $0.01<p<0.05$ implies moderate correlation and $0.1<p$ implies strong correlation.
Hence the Kolmogorov-Smirnov test  implies our model is in excellent agreement with the data.

The Kolmogorov-Smirnov test is a measure of goodness-of-fit.
We summarize this and a number of other goodness-of-fit tests as well as several variance tests in Table \ref{table-p-values}.
All show excellent agreement between model and data. Here and below we give $p$-values for the high-fit case. The low-fit values are similar.

Another approach is to calculate the $\chi^2$ for the data  set, subject to the corrections of energy per unit length differences and deformations from the ideal knot case.
We have argued that the minimum energy and minimum length of knots do not necessarily coincide.
While we expect the average  energy per unit length of a knot does not strongly depend on knot type, there is still a small correction due to this effect.
In addition, the tubes can be constricted due to being wrapped by another section of the tube or distorted by wrapping tightly  around another section of tube (as a rope wrapped tightly around a post).
We can approximate such corrections and will consider a typical example below.

\begin{figure}
\includegraphics[width=200pt]{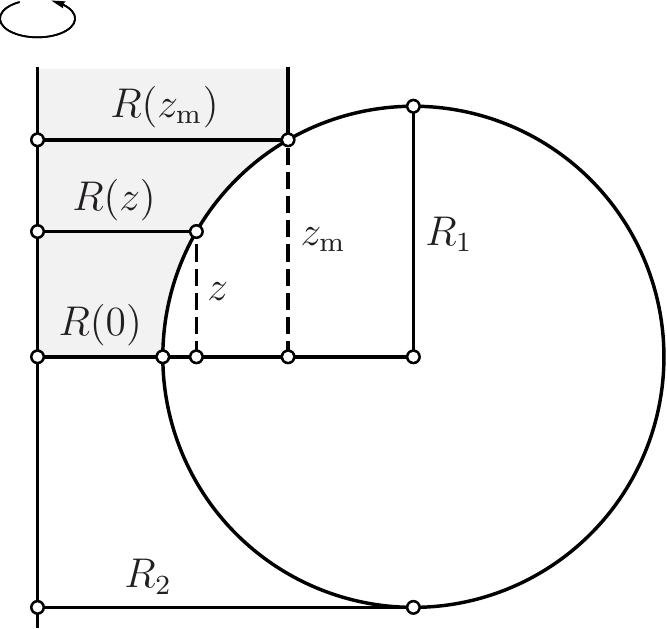}
\caption{\label{deformation} Cross section of a toroidal flux tube of minor radius $R_1$ and major radius $R_2$ constricting a cylindrical flux tube of radius $R(z)$.}
\end{figure}

We have calculated the change in energy of a pair of linked flux tubes to provide an example of corrections we should expect due to constriction, which in turn gives a contribution to the expected error in using physical tubes instead of ideal mathematical tubes to model glueballs. 
The example we consider is a  cylindrical tube along the $z$-axis encircled by a toroidal tube lying in the $xy$-plane at $z=0$.
We assume both tubes have circular cross section with that of the torus staying fixed, but that of the cylinder being of reduced radius in the region of constriction, 
\begin{align}
  R(z)=R(0)+R_1-(R_1^2-z^2)^{1/2},
  \label{}
\end{align}
where
\begin{align}
  -z_\textrm{m}\le z\le z_\textrm{m}, \quad z_\textrm{m}=[R_1^2-R^2(0)]^{1/2};
  \label{}
\end{align}
see Fig~\ref{deformation}. 
We further assume both tubes carry the same amount of flux and so set their undistorted radii equal, $R_1=R(z_\textrm{m})$.
The torus would tighten until $R_2=R_1$, except that it begins to encounter the cylindrical tube at $R_2=2R_1$.
The torus energy 
\begin{align}
\Delta W_1=W(R_2)-W(2R_1),
  \label{}
\end{align}
where the function $W$ is defined in \eqref{w}, falls as it tightens and the energy in the cylindrical tube 
\begin{align}
\Delta W_2 = \frac{\Phi_2^2}{\pi}\left[ -\frac{z_\textrm{m}}{R_2^2} +
\int_0^{z_\textrm{m}} \frac{dz}{R^2(z)} \right]
  \label{}
\end{align}
grows as it is constricted.  Stability is reached by minimizing $\Delta W_1+\Delta W_2$ with respect to $R(0)$; see Fig.~\ref{deformation}.

We have estimated the constriction for the Hopf link of two magnetic flux tubes and find a $\sim 30\%$ correction over the region of constriction, which translates into an overall $\sim 5\%$ correction to the link energy, since about $15\%$ of the Hopf link is constricted.  We expect distortion effects to be similar.
However, since all knots and links have similar corrections that modify their total energies in the same direction, we expect the spread in variation to be smaller than the correction itself. Hence a $\delta \sim 5\%$ error  on physical knot energies versus tight knot energies is not unreasonable.
An actual QCD flux tube could be more rigid due to confinement effects, so we can justify reducing the error bars to something smaller.
Even assuming QCD flux tubes are substantially more rigid than magnetic  flux tubes, we still find an acceptable value for $\chi^2$.
For example, letting $\delta_{\textrm{QCD}}=3\%$ we find   the model is in reasonable agreement with the data as seen in the figures for the fits.

In terms of the bag model~\cite{MIT-bag}, the interiors of flux tubes of tight knots correspond to the interiors of bags.
The flux in the tube is supported by current sheets on the bag boundary (surface of the tube).
Knot complexity can be reduced (or increased) by unknotting (knotting) operations~\cite{Rolfsen,Kauffman}.
In terms of flux tubes, these moves are equivalent to reconnection events~\cite{reconnection}.
Hence, a metastable glueball may decay via reconnection.
Once all topological charge is lost, metastability is lost, and the decay proceeds to completion.
Two other glueball decay processes are: flux tube or string breaking \cite{Casher:1978wy,Neuberger:1979tb,Casher:1979gw} (this favors large decay widths for configurations with long flux tube components) and quantum fluctuations that unlink flux tubes (this would tend to broaden states with short flux tube components).
Since the publication of~\cite{Buniy:2002yx}, some minor quantitative progress has been made in understanding knot flux tube decay, but these results are still insufficient to go beyond the qualitative observations made in~\cite{Buniy:2002yx}.

\begin{figure}
\includegraphics[angle=0]{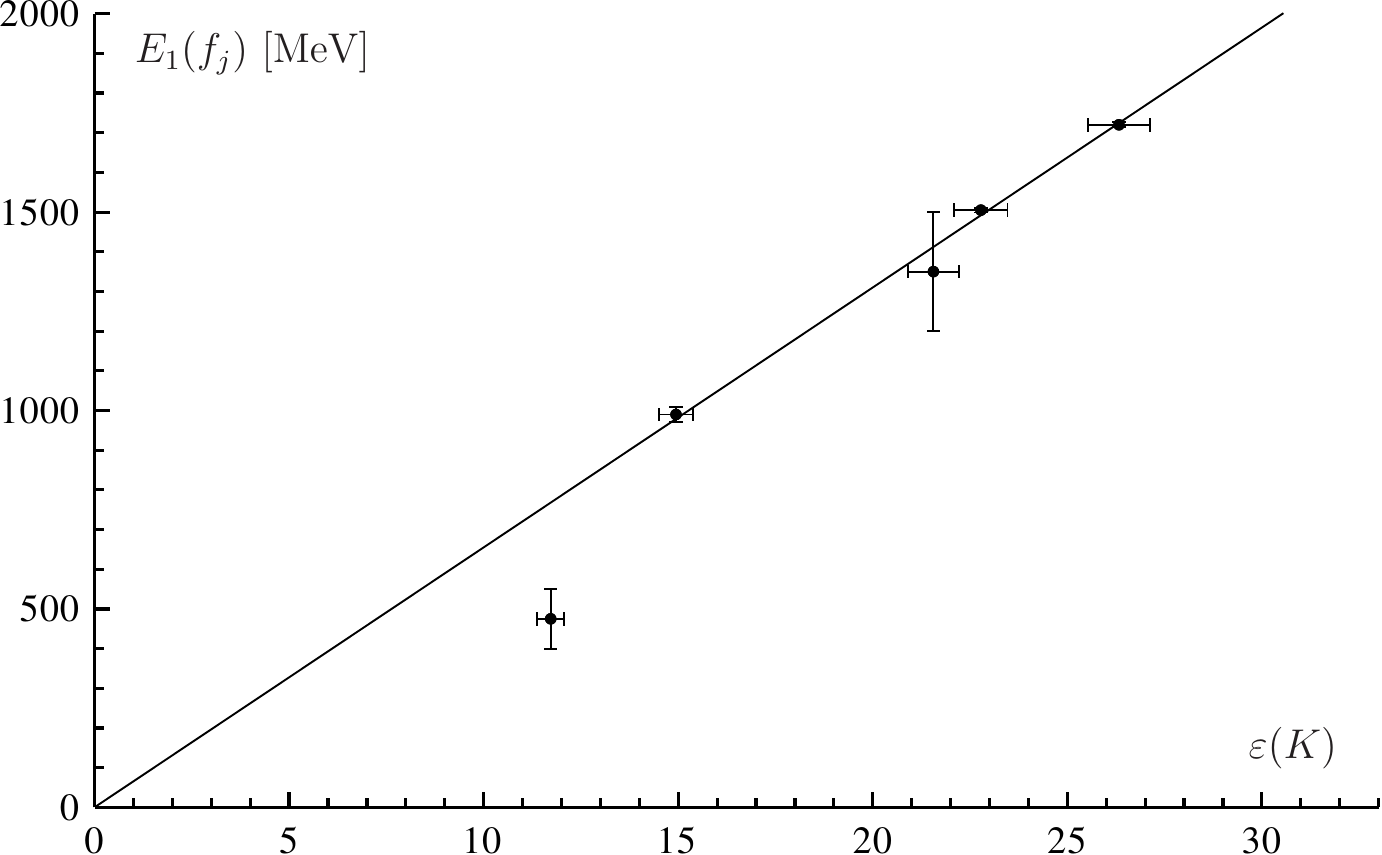}
\caption{\label{f0-fit} The $f_0$ states data is fitted to the curvature corrected knot and link data.
Errors are shown for the states and estimated to be 3\% for the knot/link energy.
Non-fitted knots and links are not shown.}
\end{figure}

Let us make one final comment about the model.
A more conservative approach is to assume only the $0^{++}$ states, i.e., the $f_0$ states, correspond to knotted/linked flux tubes.
In that case there are only five states to fit and the result is displayed in Figure \ref{f0-fit} for comparison, where the identified states are
\begin{align}
 [f_0(500),f_0(980),f_0(1370),f_0(1500),f_0(1710)]\longleftrightarrow [2^2_1,3_1,5_1,5_2,6_2].
\end{align}
The slope is essentially unchanged, the $R^2=.998$ value is roughly the same.  The masses of the predicted states can be easily gotten by rescaling the knot lengths in the table with the new slope parameter $\Lambda_{\textrm{tube}}=(65.50\pm 1.81)\,\textrm{MeV}$.
Note that the $f_0(1370)$ has been moved in the ordering to improve the fit. More data is needed to distinguish between the fit of all $f_J$ states  and the restricted $f_0$ fit. 

The  $\chi^2 = 15.62$ for the fit is not particularly good but it only takes replacing the new $f_0(500)$ values with the old $f_0(500)$ mass and error bars to get $\chi^2 = 0.56$. So if the new $f_0(500)$ mass is due to mixing and one could extract the unmixed value for the pure gluonic state it is possible that the fit would improve again.

If a sufficient number of  $f_J$ states  are found, so that they outnumber the total number of knots and links, then this would be evidence to support the restricted fit (since all short knots and links are presumed to be known).
However, this is not the case at present.

\begin{figure}
\includegraphics[angle=0]{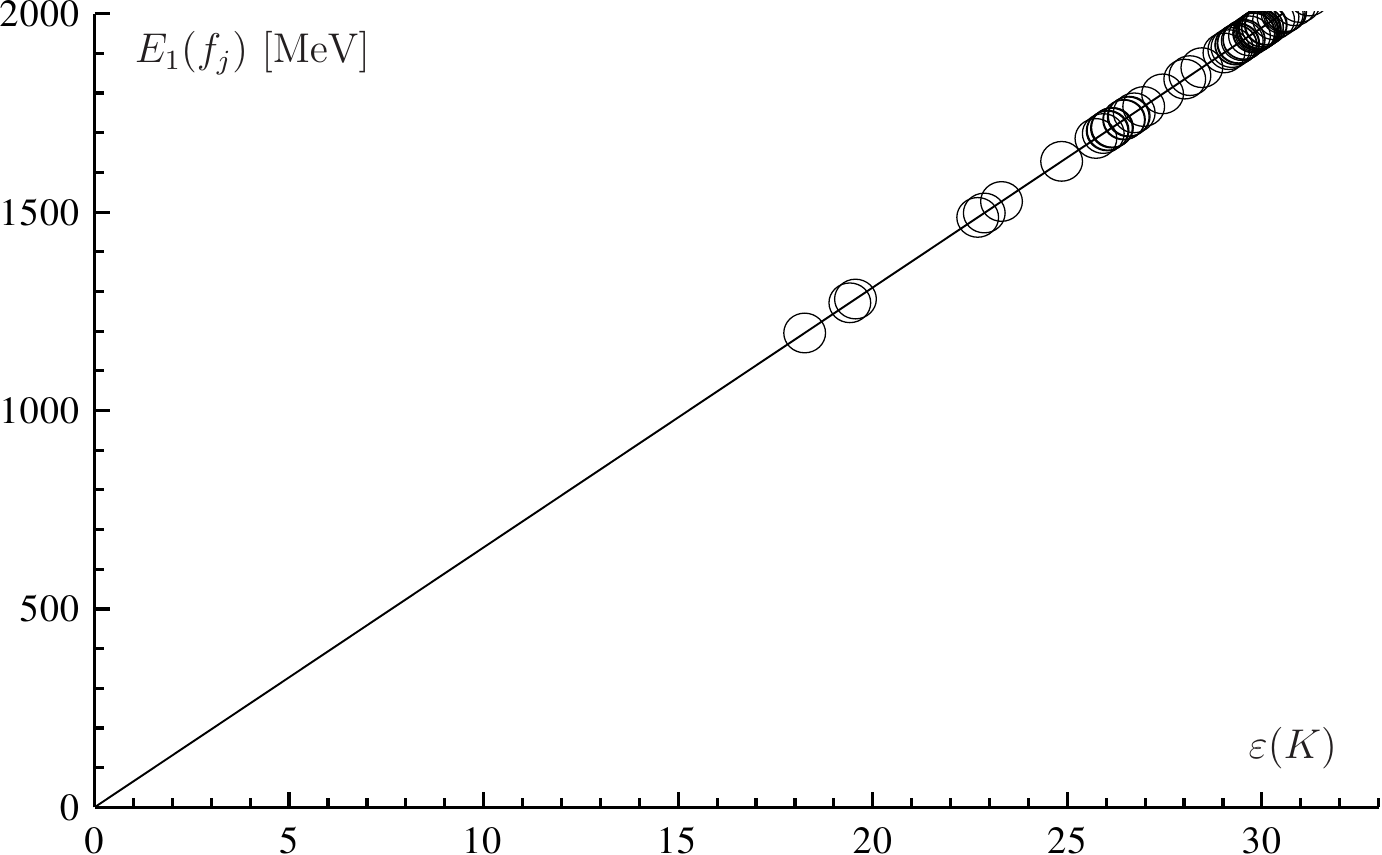}
\caption{\label{f0-fit-prediction} The $f_0$ states data is fitted to the curvature corrected knot and link data  and predicted states (open disks).}
\end{figure}

\section{Discussions and conclusions}

We have considered hadronic collisions that produce some number of baryons and mesons plus a gluonic state in the form of a closed QCD flux tube (or a set of tubes).
From an initial state, the fields in the flux tubes quickly relax to an equilibrium configuration, which is topologically equivalent to the initial state.
(We assume topological quantum numbers are conserved during this rapid process.)
The tube radius is set by the confinement scale, so to lowest order the energy of the final state depends only on the topology of the initial state and equals the length $l_K$ of the tube times the average energy per unit length, or the dimensionless knot or link length $\ve_0(K)$ times the energy scale parameter $\Lambda_{\textrm{tube}}$.
While related to $\Lambda_{\textrm{QCD}}$ by constants of order unity, $\Lambda_{\textrm{tube}}$ can be more accurately determined (see above) and hence could be a useful dimensionful parameter in studying other properties of QCD, such as scattering and hadronization processes. The relaxation proceeds through minimization of the field energy.
This process occurs via shrinking the tube length and the process halts to form a tight knot or link.
Flux conservation and energy minimization also force the fields to be homogeneous across the tube cross sections for straight tube sections and the fields fall like $1/\rho$ for curved sections as shown above. We have estimated corrections to the simple energy-length proportionality and have used them to correct and place error bars on the physical knot lengths.

Details of knot excitations would be interesting to investigate, as would other quantum corrections, but
at present we do not have a reliable way to estimate these effects, nor do we have a good way to calculate glueball decay rates.
However, we do expect high mass glueball production to be suppressed because more complicated non-trivial topological field configurations are statistically disfavored and we also expect higher mass glueballs to be relatively less stable.

On the lattice the glueball is associated with a plaquette operator
that, when operating on the vacuum, creates a closed loop of chromoelectric flux~\cite{DegrandBook}.
This  loop is a path on a square lattice and the glueball mass should
be proportional to the length of the path. Using our assumption that
topologically trivial paths are too unstable to allow measurable masses and assuming that
we are studying single flux tubes on the lattice (no links),  the first
stable closed loop on the lattice will be the trefoil. The shortest length for a trefoil on
a square lattice is 24 lattice spacings (24 tube diameters in dimensionless units),
so, without smoothing, the lattice should predict the lightest glueball mass to be a  factor of $24/\epsilon(3_1)$
larger than the value we quote in the table, i.e.,
\begin{equation}
  E_\textrm{lattice}(3_1)\sim\frac{24}{\epsilon(3_1)}E(3_1)\sim 1450 \ \textrm{MeV}.
\end{equation}
While this naive result is only a rough approximation, it could be refined and does indicate that lattice calculations of glueball masses
can come out on the high side. It is also an explanation of why our results differ from  lattice predictions.

In addition to not fitting naturally into the quark model~\cite{commentfit}, glueballs have some other characteristic signatures, including enhanced production via gluon rich channels in the central rapidity region, branching fractions incompatible with $q\bar{q}$ decay, very weak coupling to $\gamma\gamma$, and OZI suppression.
All the $f$-states have some or all of these properties.
For instance, none have substantial branching fractions to $\gamma\gamma$.
However, mixing with $q\bar{q}$ isoscalar states can obscure some of these properties.
All these observations are in qualitative agreement with the model presented here.

Our high-fit model predicts one new state at 1190 MeV, twelve states concentrated near 1700 MeV and a tower of higher mass states with the next dense concentration starting near 1900 MeV. The low-fit model makes similar prediction except that there is no new state near 1200 MeV.
We have argued that there is sufficient tension in the experimental data in these regions to allow the identification of many more states with knots and links.
A careful statistical analysis of the data of all $f$-regions to resolve hidden states is needed.
Recall we are assuming that $J$ is intrinsic angular momentum and not rotational angular momentum as we assumed in~\cite{Buniy:2002yx}.

\begin{figure}
\includegraphics[width=400pt]{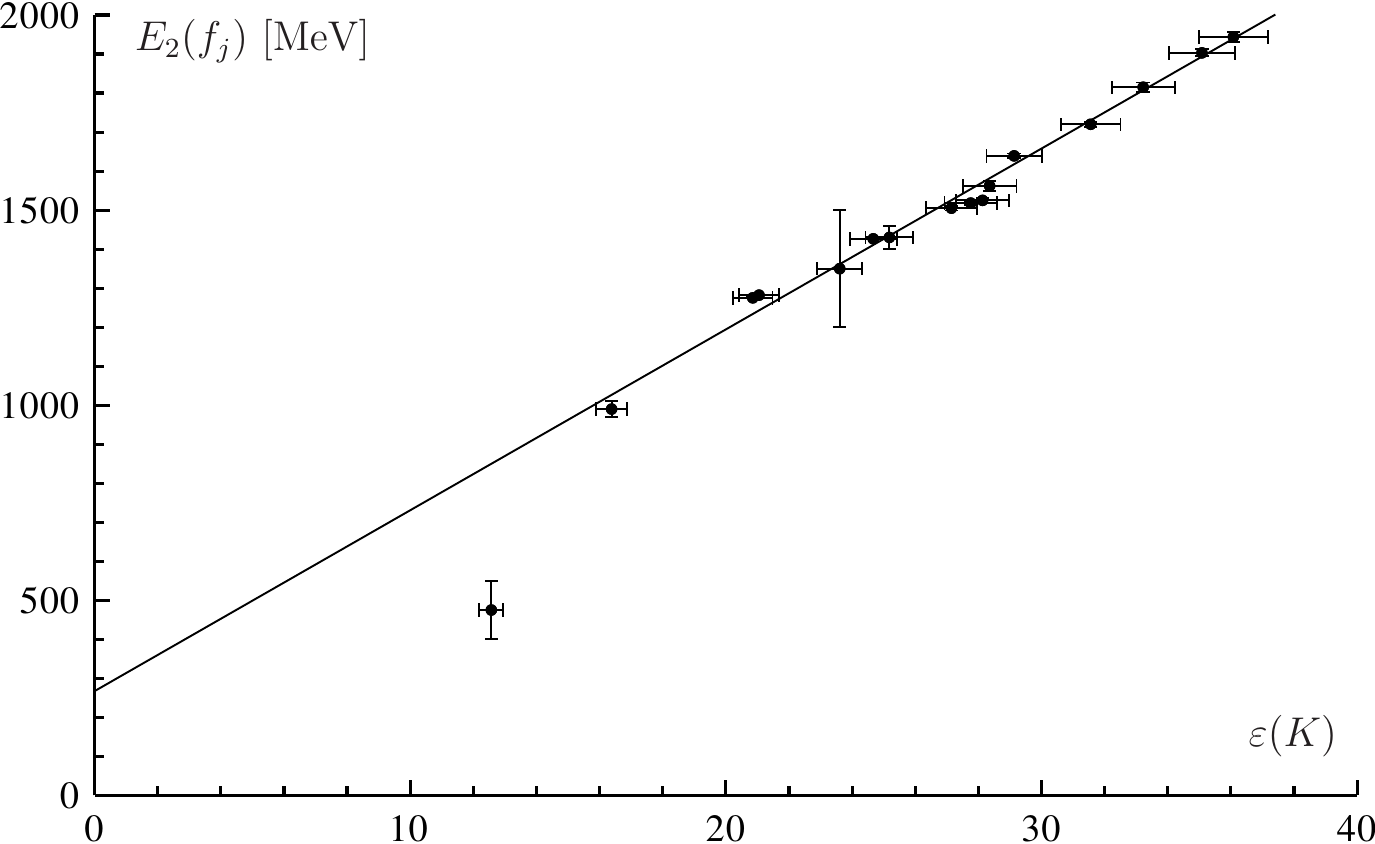}
\caption{\label{2p-fit} Uncorrected lengths:
Two parameter fit of the $f_J$ states data to the knot and link data.
Errors are shown for the states, but they are too small to be visible for the knots and links. However, a 3\% knot length error is included for reasons discussed in the text.
Non-fitted knots and links are not shown.}
\end{figure}

\begin{figure}
\includegraphics[width=400pt]{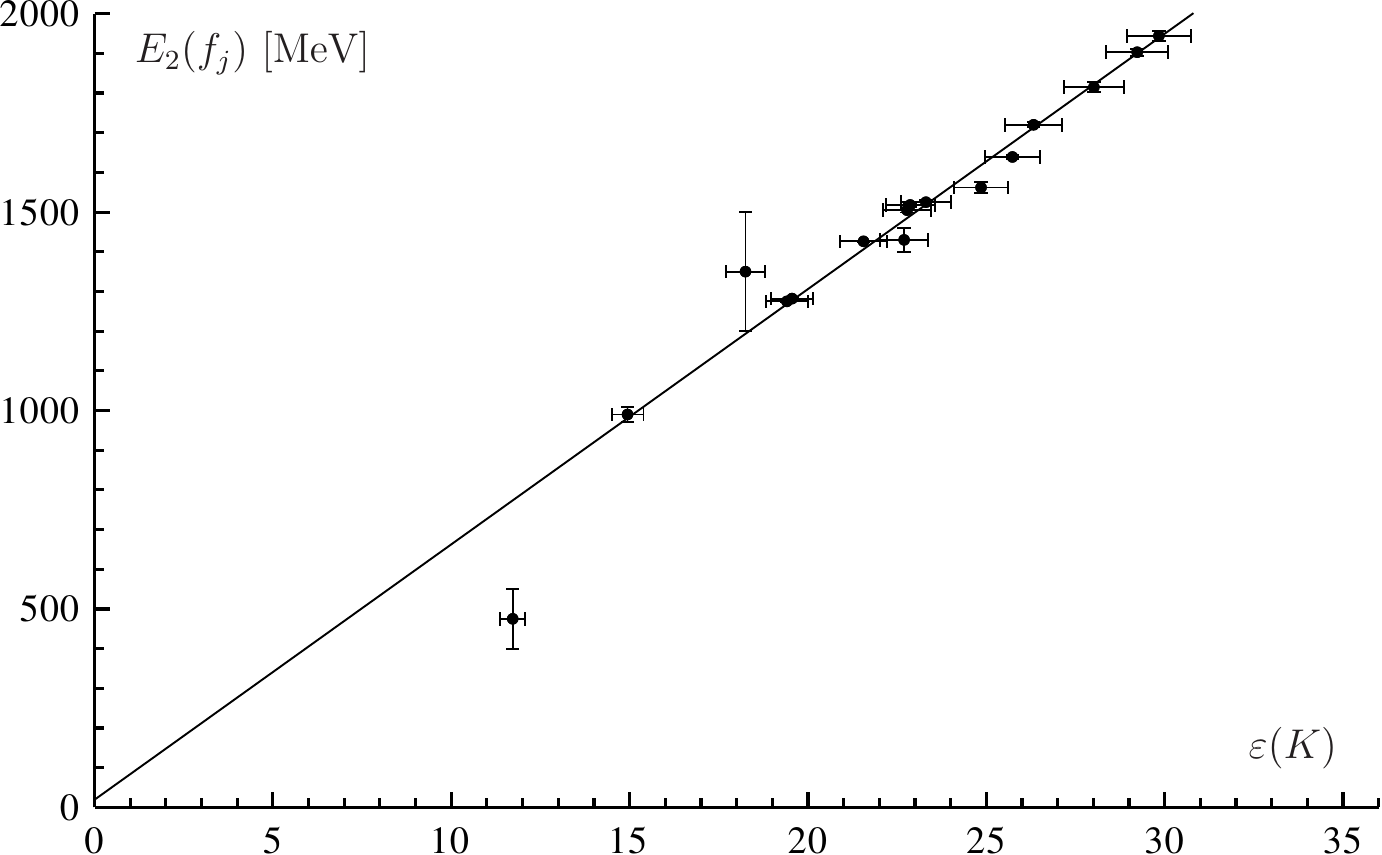}
\caption{\label{cc-low-2p-fit} Two parameter fit with curvature-corrected length.
Two parameter low $f_0(1370)$ fit  of the $f_J$ states data to the knot and link data.
Errors are shown for the $f$ states, and a 3\% error is estimated for 
knot and link lengths.
Non-fitted knots and links are not shown.}
\end{figure}

\begin{figure}
\includegraphics[width=400pt]{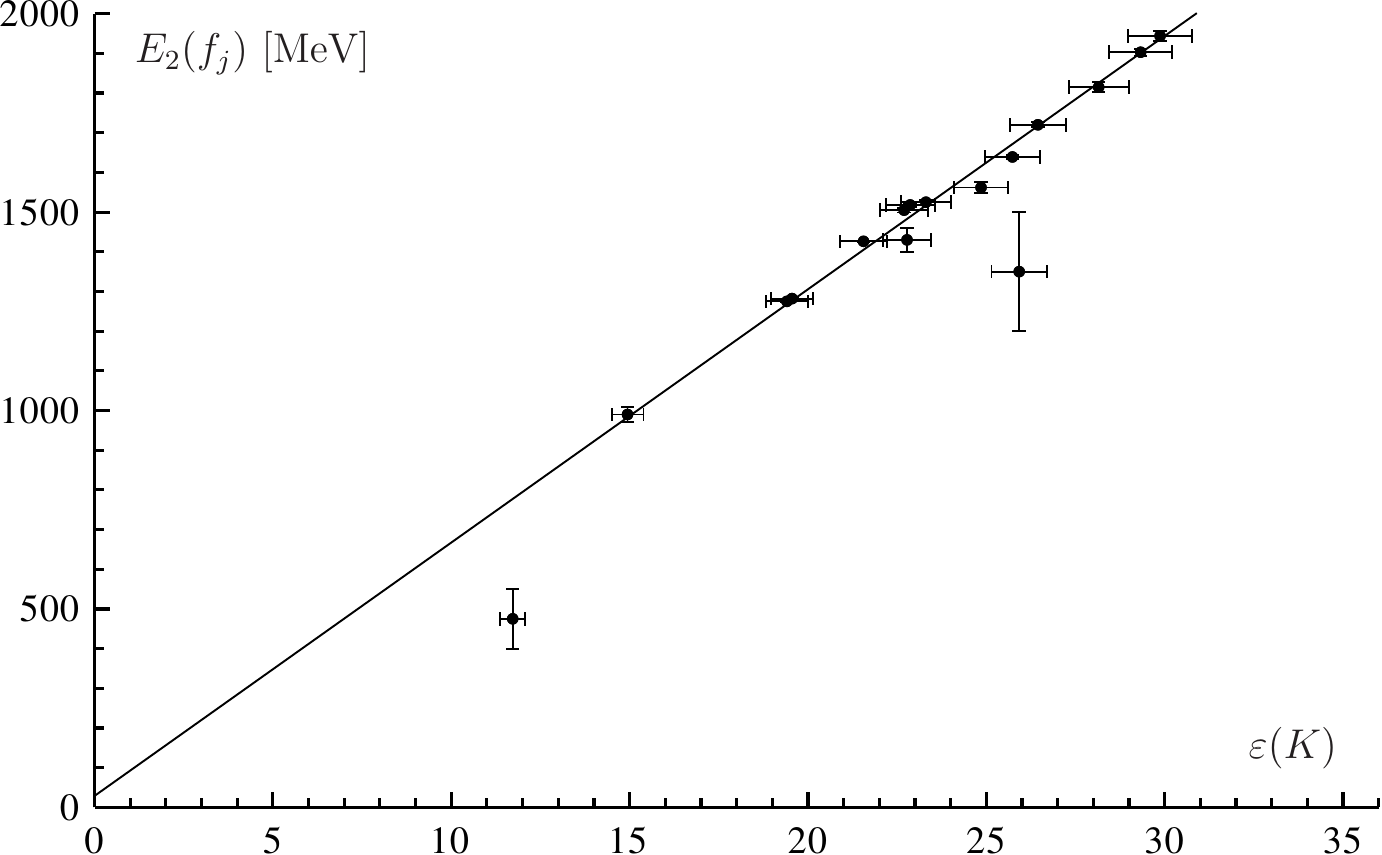}
\caption{\label{cc-high-2p-fit} Two parameter fit with curvature-corrected length.
Two parameter high $f_0(1370)$ fit of the $f_J$ states data to the knot and link data.
Errors are shown for the $f$ states and estimated for the knots and links.
Non-fitted knots and links are not shown.}
\end{figure}

As a variant example of the models  we have been considering and as a comparison, we consider a two-parameter fit to the $f_J$ data where the origin is not fixed at zero glueball mass and zero tube length. Fitting the non-curvature-corrected length data to the glueball data gives a shallower slope and an intercept at positive glueball mass; see Fig. \ref{2p-fit}. While the $\chi^2$ is somewhat better than for the non-curvature-corrected one-parameter fit, we do not have an interpretation of the non-zero intercept other than a zero length tubular bag constant which seems rather unphysical. This problem does not arise for either the high- or low-fit  two-parameter models with curvature-corrected lengths.   As seen in Figs. \ref{cc-low-2p-fit} and \ref{cc-high-2p-fit}  the intercepts in both these models are consistent with zero. This result and in addition to the improved $\chi^2$ when we use the curvature-corrected knot lengths in the one-parameter fits gives us confidence that either our  one-parameter (i.e., the slope $\Lambda_{tube}$) high or one-parameter  low curvature-corrected fits are the sufficient and best choices for a robust model.

We note that some of the states can be degenerate at lowest order due to the symmetries of some knots and links.
The trefoil comes in two versions, left- and right-handed.
Some prime knots have a $\mathbb{Z}_2\times\mathbb{Z}_2$ symmetry leading to a potential four-fold degeneracy.
The cases of links and composites are even more complicated, with a variety of higher degeneracies possible \cite{KnotSym}.
How these degeneracies can be lifted at higher order is a topic for future research, but let us make a few comments here. Since many knots and links come in more than one form, we can consider their mixings. For example, the trefoil can be left-handed $3^\textrm{L}_1$ or right-handed $3^\textrm{R}_1$. Since the trefoil can decay by changing a single crossing, changing two crossings can take $3^\textrm{L}_1$ to $3^\textrm{R}_1$. We have identified the $3_1$ with $f_0(980)$ whose width is $\sim$ 50 MeV. Assuming the decay is via reconnection or quantum tunneling, then we expect the $3^\textrm{L}_1$ and $3^\textrm{R}_1$ versions to mix at the $10^{-3}$ level by changing two crossings. This corresponds to a $\sim$1 MeV level splitting and is undetected experimentally. As the quantum state $\psi_{3_1}=\frac{1}{\sqrt{2}}(3^\textrm{L}_1 -3^\textrm{R}_1)$ has self-linking number $\textrm{SL}(\psi_{3_1})=1$ when properly normalized, while $\psi'_{3_1}=\frac{1}{\sqrt{2}}(3^\textrm{L}_1 +3^\textrm{R}_1)$ has $\textrm{SL}(\psi^{3_1})=0$, we would expect the $\psi'_{3_1}$  to be the more stable linear combination, and the proper state to be identified with the $f_0(980)$. Similar comments apply to all our identifications. Note that the Hopf link comes in only one form, so it corresponds to the single broad state $f_0(500)$.   Mixing could potentially be detected for other broad states where splitting could be a few MeV, but only when they are identified with knots or links that come in more than one type.

It would be trivial to extend our predictions to states above 2 GeV.
One just takes the knot of appropriate lengths from \cite{ridgerunner} and scales them by the dimensionful parameter $\Lambda_{\textrm{tube}}$ from the fit.
However, there is insufficient mesonic data above 2 GeV to improve or constrain our fit.
More experimental data to test the model in this region would be very welcome.

Knotted and linked magnetic fields configurations have been discussed with respect to a number of plasma phenomena including spheromaks~\cite{Spheromaks}, astrophysical~\cite{Oppermann:2010uy} and atmospheric~\cite{lightning} objects. 
Similar comments apply to Bose-Einstein condensates~\cite{Bose-Einstein}, and  various field theories. The system we consider is intrinsically different since we assume that in QCD we have confinement into tubes and then the tubes get knotted and linked, as opposed to finding knotted or linked fields that then may or may not get confined. In our case we expect energy proportional to tube length (with corrections as discussed in this paper), where the later case not involving initial confinement would not necessarily be expected to have a simple length-energy relationship.
The energies and sizes of these classical solitons are sometimes difficult to quantify since they depend on parameters of the plasma, including temperature, pressure, density, ionic content, etc.
However, it was argued in~\cite{Buniy:2002yx} that well-defined topological soliton energies can be identified for vacuum QCD or any vacuum quantum-flux-tube system.
Hence we emphasize that in all systems which support quantum flux tubes, including those occuring in media, from the quark-gluon plasma to superconductors, the energy spectrum of knot and link solitons will be universal up to a scaling for fixed values of parameters~\cite{Buniy:2004zw,Buniy:2004zy}.

To conclude, we have given two interpretations of the $f$-state data with a model of knotted chromoelectric flux tubes in QCD. The first possibility is where the $f(1370)$ is identified with the $5_1$ knot which results in a prediction of a new state around 1190 MeV which is identified with the $4^2_1$ link. The other possibility---which also gives our best fit---is to identify the $f(1370)$ with the $4^2_1$ link to give a one-to-one matching of all the first twelve $f_J$ states with the first twelve knots and links. Experiments could help to resolve which of these two possibilities is the correct choice. 

Finally we should point out that there is considerable amount of tension in the $f_J$ data, as indicate by the $\chi^2$s of the individual states quoted by the PDG. We would not expect our fits to be better than the fits of the data on which they are based. 

\begin{acknowledgments}
We have benefited from discussions and email correspondences with many colleagues over the last several years.
We especially thank James Bjorken, Hai-Yang Cheng, and Ryan Rohm for useful comments.
We also thank the Isaac Newton institute for hospitality while this work was being completed.
The work of TWK was supported by U.S. DoE grant number DE-FG05-85ER40226, and that of EJR
by NSF DMS \#1115722. Many of the knot and link lengths were calculated using Vanderbilt's Advanced Computing Center for Research and Education (ACCRE).
\end{acknowledgments}

\begingroup
\squeezetable
\begin{table}[htb]
\caption{\label{table-lengths1}Comparison of the glueball mass spectrum and fit energies for $\ve_0(K)$ less than $\sim 36$. $E_1(K)$ and $E_2(K)$ are for one and two parameter high $f_0(1370)$ fits.}
\begin{ruledtabular}
\begin{tabular}{cccccc}
State        & Mass            & $K\footnotemark[1]$ & $\ve_0(K)\footnotemark[2]$ & $E_1(K) $ & $E_2(K)$ \\ \hline
$f_0(500)$   & $475\pm 75$     & $2^2_1$         & $12.571\footnotemark[3]$                         & $718 $ & $850 $ \\
$f_0(980)$   & $990\pm 20$     & $3_1$           & $16.381$                                         & $936 $ & $1026$ \\
             &                 & $4^2_1$         & $20.011$                                         & $1143$ & $1195$ \\
$f_2(1270)$  & $1275.1\pm 1.2$ & $2^2_1\#2^2_1$  & $20.853\footnotemark[3]$                         & $1192$ & $1234$ \\
$f_1(1285)$  & $1282.1\pm 0.6$ & $4_1$           & $21.051$                                         & $1203$ & $1243$ \\
$f_0(1370)$  & $1350\pm 150$   & $5_1$           & $23.608$                                         & $1349$ & $1361$ \\ 
$f_1(1420)$  & $1426.4\pm 0.9$ & $2^2_1\#3_1$    & $24.671$                                         & $1410$ & $1411$ \\
$f_2(1430)$  & $\approx 1430$  & $5_2$           & $24.745$                                         & $1414$ & $1414$ \\ 
$f_0(1500)$  & $1505\pm 6$     & $5^2_1$         & $24.893$                                         & $1422$ & $1421$ \\
$f_1(1510)$  & $1518\pm 5$     & $6^3_3$         & $25.181$                                         & $1439$ & $1434$ \\ 
$f'_2(1525)$ & $1525\pm 5$     & $6^2_1$         & $27.146$                                         & $1551$ & $1525$ \\
$f_2(1565)$  & $1562\pm 13$    & $7^2_7$         & $27.760$                                         & $1586$ & $1554$ \\
$f_2(1640)$  & $1639\pm 6$     & $(2^2_1\#2^2_1\#2^2_1)_{\textrm{kc}}$ & $28.133\footnotemark[3]$   & $1608$ & $1571$ \\
             &                 & $2^2_1\#4^2_1$  & $28.311$                                         & $1618$ & $1579$ \\
             &                 & $6^2_2$         & $28.356$                                         & $1620$ & $1582$ \\
             &                 & $6_1$           & $28.364$                                         & $1621$ & $1582$ \\
             &                 & $6_2$           & $28.522$                                         & $1630$ & $1589$ \\ 
	     &     	       & $3^1_m\#3^1_m$  & $28.537$                                         & $1631$ & $1590$ \\
	     &     	       & $2^2_1\#4_1$    & $28.742$                                         & $1642$ & $1599$ \\
	     &     	       & $7^2_8$         & $28.886$                                         & $1651$ & $1606$ \\
	     &     	       & $6^3_1$         & $28.914$                                         & $1652$ & $1607$ \\
	     &    	       & $6_3$           & $28.929$                                         & $1653$ & $1608$ \\
	     &    	       & $6^3_2$         & $29.006$                                         & $1657$ & $1612$ \\
	     &    	       & $6^2_3$         & $29.057$                                         & $1660$ & $1614$ \\
	     &    	       & $(2^2_1\# 2^2_1\# 2^2_1)_{\textrm{lc}}$ & $29.133\footnotemark[3]$ & $1665$ & $1618$ \\
$[f_0(1710)]_1$  & $[1720\pm 6]_1$  & $8^3_7$    & $30.297$                                         & $1731$ & $1672$ \\
             &                 & $8_{19}$        & $30.502$                                         & $1743$ & $1681$ \\
             &                 & $7_1$           & $30.715$                                         & $1755$ & $1691$ \\
$[f_0(1710)]_2$ & $[1720\pm 6]_2$ & $8_{20}$     & $31.557$                                         & $1803$ & $1730$ \\
$[f_2(1810)]_1$ & $[1815\pm 12]_1$ & $2^2_1\#5_1$ & $31.908$                                         & $1823$ & $1746$ \\
             &                 & $7_3$           & $31.975$                                         & $1827$ & $1749$ \\
             &                 & $8^2_{15}$      & $32.093$                                         & $1834$ & $1755$ \\
             &                 & $7_2$           & $32.122$                                         & $1836$ & $1756$ \\
             &                 & $7^2_1$         & $32.129$                                         & $1836$ & $1756$ \\
	     &                 & $7_4$           & $32.146$                                         & $1837$ & $1757$ \\
             &                 & $3_{1m}\#4^2_1$ & $32.189$                                         & $1839$ & $1759$ \\
	     &                 & $8^3_8$         & $32.514$                                         & $1858$ & $1774$ \\
	     &                 & $7^2_2$         & $32.520$                                         & $1858$ & $1775$ \\
	     &                 & $7^2_4$         & $32.542$                                         & $1860$ & $1776$ \\
\end{tabular}
\end{ruledtabular}			
\end{table}
\endgroup

\begingroup
\squeezetable
\begin{table}[htb]
\caption{\label{table-lengths2}Comparison of the glueball mass spectrum and  fit energies for $\ve(K)$ less than $\sim 36$.}
\begin{ruledtabular}
\begin{tabular}{cccccc}
State & Mass        & $K\footnotemark[1]$ & $\ve(K)\footnotemark[2]$ & $E_1(K)\footnotemark[3]$ & $E_2(K)\footnotemark[4]$ \\ \hline
& & $3_1\# 4_1$     & $32.636$                   & $1865$  & $1780$ \\
& & $7_5$           & $32.641$                   & $1865$  & $1780$ \\
& & $2^2_1\# 5^2_1$ & $32.659$                   & $1866$  & $1781$ \\
& & $7^2_3$         & $32.675$                   & $1867$  & $1782$ \\
& & $8^3_{10}$      & $32.735$                   & $1871$  & $1785$ \\
& & $8^{21}$        & $32.771$                   & $1873$  & $1786$ \\
& & $7_7$           & $32.816$                   & $1875$  & $1788$ \\
& & $7_6$           & $32.857$                   & $1878$  & $1790$ \\
& & $7^2_5$         & $32.873$                   & $1878$  & $1791$ \\
& & $7^3_1$         & $32.911$                   & $1881$  & $1793$ \\
& & $2^2_1\# 2^2_1\# 3_{1\textrm{B}}$ & $32.959$ & $1883$  & $1795$ \\
& & $9^2_{49}$      & $33.028$                   & $1887$  & $1798$ \\
& & $2^2_1\# 5_2$   & $33.041$                   & $1888$  & $1799$ \\
& & $9^2_{43}$      & $33.134$                   & $1893$  & $1803$ \\
& & $7^2_6$         & $33.167$                   & $1895$  & $1805$ \\
& & $8^2_{16}$      & $33.221$                   & $1898$  & $1807$ \\
$[f_2(1910)]_1,[f_2(1810)]_2$ & $[1903\pm 9]_1,[1815\pm 12]_2$ & $8^3_9$    & $33.359$                   & $1906$  & $1813$ \\
& & $8^4_2$         & $33.711$                   & $1926$  & $1830$ \\
$[f_2(1950)]_1$ & $[1944\pm 12]_1$ & $9^2_{53}$      & $34.008$                   & $1943$  & $1844$ \\
& & $8^2_1$         & $34.214$                   & $1955$  & $1853$ \\
& & $9_{46}$        & $34.319$                   & $1961$  & $1858$ \\
& & $9^2_{50}$      & $34.350$                   & $1963$  & $1859$ \\
& & $9^2_{61}$      & $34.689$                   & $1982$  & $1875$ \\
& & $9_{42}$        & $34.755$                   & $1986$  & $1878$ \\
& & $9^2_{47}$      & $35.083$                   & $2005$  & $1893$ \\
$[f_2(1910)]_2$ & $[1903\pm 9]_2$ & $9^2_{51}$      & $35.276$                   & $2016$  & $1902$ \\
& & $(2^2_1\# 2^2_1\#2^2_1\#2^2_1)_{kc}$        & $35.416$                   &    $2028$&$1912$   \\
& & $9^2_{54}$        & $35.516$                   & $2029$  & $1913$ \\
& & $8^2_2$	    & $35.526$                   & $2030$  & $1914$ \\
& & $10_{124}$	    & $35.555$                   & $2032$  & $1915$ \\
& & $8_3$           & $35.745$                   & $2043$  & $1924$ \\
& & $9_{44}$	    & $35.844$                   & $2048$  & $1929$ \\
& & $9^2_{56}$	    & $35.895$                   & $2051$  & $1931$ \\
$[f_2(1950)]_2$ & $[1944\pm 12]_2$ & $8_5$	    & $36.080$                   & $2062$  & $1940$ \\
\end{tabular}
\end{ruledtabular}
\footnotetext[1]{Notation $n^l_k$ means a link of $l$ components with $n$ crossings of $k^\textrm{th}$ type, see e.g., \protect\cite{Rolfsen}.
$K\#K'$ stands for the connected sum of $K$ and $K'$, and $(K)_\textrm{m}$ is the mirror image of $K$.} 
\footnotetext[2]{All values of the ropelength $\ve_0(K)$ are  from \protect\cite{ridgerunner}, except for composite links that were calculated separately but also with \texttt{Ridgerunner}. The dimensionless length convention agrees with \cite{Buniy:2002yx} and is a factor of 2 smaller than that in \cite{ridgerunner}.}
\footnotetext[3]{$E(K)$ is the value of the fitted mass corresponding to the PDG or predicted mass obtained from $\ve(K)$.}
\footnotetext[4] {Dimensionless curvature corrected knot energy.}
\footnotetext[5]{Exact values: $\epsilon_0(2^2_1)=4\pi$ (Hopf link), $\epsilon_0(2^2_1\#2^2_1)=6\pi+2$ (chain of three links), $\epsilon_0((2^2_1\#2^2_1\#2^2_1)_{\textrm{kc}})=8\pi+3$ (key chain link with three keys), $\epsilon_0((2^2_1\# 2^2_1\# 2^2_1)_{\textrm{lc}})=8\pi+4$ (linear chain with four links), and $\epsilon_0((2^2_1\#2^2_1\#2^2_1\#2^2_1)_{\textrm{kc}})=10\pi+4$ (key chain link with four keys).}
\end{table}
\endgroup

\begingroup
\squeezetable
\begin{table}[htb]
\caption{\label{table-cclengths1} Low $f_0(1370)$ curvature corrected fit: Comparison of the glueball mass spectrum and fit energies for $\ve(K)$ less than $\sim 32$.}
\begin{ruledtabular}
\begin{tabular}{ccccc}
State        & Mass                & $K\footnotemark[1]$ & $\ve(K)\footnotemark[2]$ & $E_1(K)\footnotemark[3]$ \\ \hline
$f_0(500)$   & $475\pm 75$         &  $2^2_1	                          $ & $11.724$ & $764 $\\ 
$f_0(980)$   & $990\pm 20$         &  $3_1	                          $ & $14.943$ & $974 $\\ 
$f_0(1370)$  & $1350\pm 150$      &  $4^2_1	                          $ & $18.250$ & $1189$\\ 
$f_2(1270)$  & $1275.1\pm 1.2$     &  $4_1	                          $ & $19.411$ & $1265$\\ 
$f_1(1285)$  & $1282.1\pm 0.6$     &  $2^2_1\# 2^2_1                      $ & $19.556$ & $1274$\\ 
$f_1(1420)$  & $1426.4\pm 0.9$    &  $5_1	                          $ & $21.559$ & $1405$\\ 
$f_2(1430)$  & $\approx 1430$     &  $2^2_1\# 3_1                         $ & $22.697$ & $1479$\\ 
$f_0(1500)$  & $1505\pm 6$        &  $5_2	                          $ & $22.779$ & $1484$\\ 
$f_1(1510)$  & $1518\pm 5$        &  $5^2_1	                          $ & $22.866$ & $1490$\\ 
$f'_2(1525)$ & $1525\pm 5$        &  $6^3_3	                          $ & $23.309$ & $1519$\\ 
$f_2(1565)$  & $1562\pm 13$       &  $6^2_1	                          $ & $24.854$ & $1619$\\ 
 $f_2(1640)$  & $1639\pm 6$       &  $7^2_7	                          $ & $25.735$ & $1677$\\ 
              &                   &  $6^2_2	                          $ & $25.924$ & $1689$\\ 
             &                     &  $6_1	                          $ & $26.025$ & $1696$\\ 
             &                     &  $2^2_1\# 4^2_1                      $ & $26.046$ & $1697$\\ 
             &                     &  $3_1\# 3_{1m}                       $ & $26.135$ & $1703$\\ 
             &                     &  $3_1\# 3_1                          $ & $26.151$ & $1704$\\ 
	     &     	           &  $6_2	                          $ & $26.158$ & $1704$\\ 
$f_0(1710)$ & $1720\pm 6$          &  $6^3_1	                          $ & $26.327$ & $1715$\\ 
	     &     	           &  $(2^2_1\#2^2_1\#2^2_1)_{\textrm{kc}}$ & $26.449$ & $1723$\\ 
	     &     	           &  $2^2_1\# 4_1                        $ & $26.466$ & $1724$\\ 
	     &    	           &  $6_3	                          $ & $26.567$ & $1731$\\ 
	     &    	           &  $6^2_3	                          $ & $26.590$ & $1733$\\ 
	     &    	           &  $7^2_8	                          $ & $26.720$ & $1741$\\ 
	     &    	           &  $6^3_2	                          $ & $26.963$ & $1757$\\ 
	     &    	           &  $(2^2_1\#2^2_1\#2^2_1)_{\textrm{lc}}$ & $27.449$ & $1788$\\ 
$f_2(1810)$ & $1815\pm 12$         &  $7_1                                $ & $28.018$ & $1826$\\ 
             &                     &  $8^3_7	                          $ & $28.152$ & $1834$\\ 
             &                     &  $8_{19}	                          $ & $28.458$ & $1854$\\ 
             &                     &  $7_3	                          $ & $29.025$ & $1891$\\ 
             &                     &  $8_{20}	                          $ & $29.151$ & $1899$\\ 
             &                     &  $7^2_1	                          $ & $29.231$ & $1905$\\ 
             &                     &  $7_2	                          $ & $29.330$ & $1911$\\ 
             &                     &  $2^2_1\# 5_1                        $ & $29.339$ & $1912$\\ 
	     &                     &  $7_4	                          $ & $29.385$ & $1915$\\ 
             &                     &  $3_{1m}\# 4^2_1                     $ & $29.402$ & $1916$\\ 
             &                     &  $8^2_{15}	                          $ & $29.496$ & $1922$\\ 
	     &                     &  $3_1\# 4^2_1                        $ & $29.536$ & $1924$\\ 
	     &                     &  $7^2_4	                          $ & $29.544$ & $1925$\\ 
\end{tabular}
\end{ruledtabular}			
\end{table}
\endgroup

\begingroup
\squeezetable
\begin{table}[htb]
\caption{\label{table-cclengths2}Comparison of the glueball mass spectrum and  fit energies for $\ve(K)$ less than $\sim 32$.}
\begin{ruledtabular}
\begin{tabular}{ccccc}
State & Mass        & $K\footnotemark[1]$ & $\ve(K)\footnotemark[2]$ & $E_1(K)\footnotemark[3]$ \\ \hline
$ $ & $ $ & $2^2_1\# 2^2_1\# 3_{1A} $ & $29.682$        & $1934 $ \\
$ $ & $ $ & $3_1\# 4_1	            $ & $29.790$        & $1941 $ \\
$ $ & $ $ & $7_5	            $ & $29.806$        & $1942 $ \\
$f_2(1950) $ & $1944\pm 12 $ & $2^2_1\# 5_2$ & $29.840$ & $1944 $ \\
$ $ & $ $ & $7^2_3	            $ & $29.873$        & $1946 $ \\
$ $ & $ $ & $7_6	            $ & $29.894$        & $1948 $ \\
$ $ & $ $ & $7^2_2	            $ & $29.895$        & $1948 $ \\
$ $ & $ $ & $2^2_1\# 5^2_{1m}	    $ & $29.929$        & $1950 $ \\
$ $ & $ $ & $2^2_1\# 5^2_{1A}       $ & $29.952$        & $1952 $ \\
$ $ & $ $ & $8^3_8	            $ & $29.957$        & $1952 $ \\
$ $ & $ $ & $7^2_5	            $ & $30.015$        & $1956 $ \\
$ $ & $ $ & $8_{21}	            $ & $30.017$        & $1956 $ \\
$ $ & $ $ & $7_7	            $ & $30.092$        & $1961 $ \\
$ $ & $ $ & $7^3_1	            $ & $30.112$        & $1962 $ \\
$ $ & $ $ & $7^2_6	            $ & $30.302$        & $1974 $ \\
$ $ & $ $ & $9^2_{43}	            $ & $30.416$        & $1982 $ \\
$ $ & $ $ & $8^2_{16}	            $ & $30.525$        & $1989 $ \\
$ $ & $ $ & $8^3_9	            $ & $30.571$        & $1992 $ \\
$ $ & $ $ & $8^3_{10}	            $ & $30.605$        & $1994 $ \\
$ $ & $ $ & $2^2_1\# 2^2_1\# 3_{1B} $ & $30.611$        & $1995 $ \\
$ $ & $ $ & $9^2_{49}	            $ & $30.839$        & $2009 $ \\
$ $ & $ $ & $8^4_2	            $ & $30.967$        & $2018 $ \\
$ $ & $ $ & $8^2_1	            $ & $31.214$        & $2034 $ \\
$ $ & $ $ & $8^4_3	            $ & $31.473$        & $2051 $ \\
$ $ & $ $ & $9_{46}	            $ & $31.513$        & $2053 $ \\
$ $ & $ $ & $9^2_{50}	            $ & $31.521$        & $2054 $ \\
$ $ & $ $ & $9^2_{53}	            $ & $31.776$        & $2070 $ \\
$ $ & $ $ & $9^2_{61}	            $ & $31.909$        & $2079 $ \\
$ $ & $ $ & $9^2_{47}	            $ & $31.916$        & $2080 $ \\
$ $ & $ $ & $9_{42}	            $ & $31.950$        & $2082 $ \\
\end{tabular}
\end{ruledtabular}
\footnotetext[1]{Notation $n^l_k$ means a link of $l$ components with $n$ crossings of $k^\textrm{th}$ type, see e.g., \protect\cite{Rolfsen}.
$K\#K'$ stands for the connected sum of $K$ and $K'$, and $(K)_\textrm{m}$ is the mirror image of $K$.} 
\footnotetext[2]{All values of the curvature corrected knot energies $\phi(K)$ are modified from \protect\cite{ridgerunner}, except for composite links that were calculated separately but also with \texttt{Ridgerunner}. The dimensionless length convention agrees with \cite{Buniy:2002yx} and is a factor of 2 smaller than that in \cite{ridgerunner}.}
\footnotetext[3]{$E_1(K)$ is the value of the fitted mass corresponding to the PDG or predicted mass obtained from $\ve(K)$.}
\footnotetext[4] {Dimensionless curvature corrected knot energy.}
\footnotetext[5]{Exact values: $\epsilon_0(2^2_1)=4\pi$ (Hopf link), $\epsilon_0(2^2_1\#2^2_1)=6\pi+2$ (chain of three links), $\epsilon_0((2^2_1\#2^2_1\#2^2_1)_{\textrm{kc}})=8\pi+3$ (key chain link with three keys), $\epsilon_0((2^2_1\# 2^2_1\# 2^2_1)_{\textrm{lc}})=8\pi+4$ (linear chain with four links), and $\epsilon_0((2^2_1\#2^2_1\#2^2_1\#2^2_1)_{\textrm{kc}})=10\pi+4$ (key chain link with four keys).}
\end{table}
\endgroup

\begingroup
\squeezetable
\begin{table}[htb]
\caption{\label{table-p-values}Statistical tests of the model.
Recalling that $p$ is bounded $0\le p\le 1$ and $p<0.01$ implies poor correlation, $0.01<p<0.05$ implies moderate correlation and $0.1<p$ implies strong correlation, we see that all these tests strongly support the model.
}
\begin{ruledtabular}
\begin{tabular}{llll}
Goodness-of-fit test & $p$-value & Variance test & $p$-value \\ \hline
Pearson $\chi^2$ & 0.66 & Brown-Forsythe & 0.74\\
Kolmogorov-Smirnov & $0.95$ & Fisher Ratio & 0.69\\
Cramer-von Mises & 0.96 & Levene & 0.74\\
Anderson-Darling &0.97 & Siegel-Tukey & 0.82\\
Kuiper & 0.99\\
Watson U Square & 0.90
\end{tabular}
\end{ruledtabular}
\end{table}
\endgroup

\begingroup
\squeezetable
\begin{table}[htb]
\caption{\label{fitparameters}Fit parameters for the model at 95\% CL.
We collect the results for the various choices of one and to parameter fits so raw length fits can be compared with curvature corrected fits.
}
\begin{ruledtabular}
\begin{tabular}{ccccccc}
Fitting Parameter & $1p$-length & $2p$-length & $1p$-cc-high &$2p$-cc-high&$1p$-cc-low&$2p$-cc-low\\ \hline
 $\chi^2$  & 83.9 & 45.4 & 33.3 & 33.1&28.3&28.4\\
$\Lambda_{tube}$ & $57.14$ $\pm$ $0.53$ & $46.36$ $\pm$ $2.64$& $65.06$ $\pm$ $0.61$&$63.83$ $\pm$ $3.57$&$65.16$ $\pm$ $0.61$&$64.34$ $\pm$ $3.59$\\
Intercept & $0$ & $267.11$ $\pm$ $69.55$&$0$&$28.74$ $\pm $ $82.18$&$0$&$19.16$ $ \pm$ $ 82.51$\\
$R^2$&0.998 & .967 & .999&.977&.999&.980\\
\end{tabular}
\end{ruledtabular}
\end{table}
\endgroup

\begingroup
\squeezetable
\begin{table}[htb]
\caption{\label{table-cclengths1H} High $f_0(1370)$ curvature corrected fit: Comparison of the glueball mass spectrum and fit energies for $\ve(K)$ less than $\sim 32$. Except for the $f(4^2_1)$, this table contains only the fitted PDG states. Predictions for other states can be gotten by multiplying the knot energy by the appropriate value of $\Lambda_\textrm{tube}$.}
\begin{ruledtabular}
\begin{tabular}{ccccc}
State        & Mass                & $K$ & $\ve(K)$ & $E_1(K)$ \\ \hline
$f_0(500)$   & $475\pm 75$         &  $2^2_1	                          $ & $11.724$ & $763 $\\ 
$f_0(980)$   & $990\pm 20$         &  $3_1	                          $ & $14.943$ & $972 $\\ 
 $f(4^2_1)$  &    &  $4^2_1	                          $ & $18.250$ & $1187$\\ 
$f_2(1270)$  & $1275.1\pm 1.2$     &  $4_1	                          $ & $19.411$ & $1263$\\ 
$f_1(1285)$  & $1282.1\pm 0.6$     &  $2^2_1\# 2^2_1                      $ & $19.556$ & $1272$\\ 
$f_1(1420)$  & $1426.4\pm 0.9$    &  $5_1	                          $ & $21.559$ & $1403$\\ 
$f_0(1500)$  & $1505\pm 6$        & $2^2_1\# 3_1                         $ & $22.697$ & $1477$\\ 
$f_2(1430)$  & $\approx 1430$     &  $5_2	                          $ & $22.779$ & $1482$\\ 
$f_1(1510)$  & $1518\pm 5$        &  $5^2_1	                          $ & $22.866$ & $1488$\\ 
$f'_2(1525)$ & $1525\pm 5$        &  $6^3_3	                          $ & $23.309$ & $1516$\\ 
$f_2(1565)$  & $1562\pm 13$       &  $6^2_1	                          $ & $24.854$ & $1617$\\ 
 $f_2(1640)$  & $1639\pm 6$       &  $7^2_7	                          $ & $25.735$ & $1674$\\ 
 $f_0(1370)$               &      $1350\pm 150$                &  $6^2_2	                          $ & $25.924$ & $1687$\\ 
 $f_0(1710)$ & $1720\pm 6$          &  $(2^2_1\#2^2_1\#2^2_1)_{\textrm{kc}}$ & $26.449$ & $1721$\\
 $f_2(1810)$ & $1815\pm 12$         &  $7_1                                $ & $28.018$ & $1823$\\ 
 $f_2(1910)$ & $1903\pm 9$ & $7_2$    & $29.330$                   & $1908$  \\ 
 $f_2(1950) $ & $1944\pm 12 $ & $2^2_1\# 5_2$ & $29.840$ & $1941 $ \\
\end{tabular}
\end{ruledtabular}			
\end{table}
\endgroup


\begin{thebibliography}{99}

\bibitem{LK}
Sir William Thomson (Lord Kelvin), 
Proc. Royal Soc. of Edinburgh, Vol. VI, 94 (1867).

\bibitem{PDG}
 Particle Data Group (PDG),
  K. Nakamura et al., Journal of Physics G {\bf 37}, 075021 (2010).

\bibitem{Nielsen-Olesen}
H.~B. Nielsen and P.~Olesen,
Nucl. Phys., B {\bf 61}, 45 (1973).

\bibitem{Swanson} E.~S.~Swanson, {\em 7th International Conference on
Hadron Spectroscopy (Hadron 97)}, Upton, NY, AIP Conf.\ Proc.\ {\bf
432}, 471 (1997).

\bibitem{West} For further discussion of glueballs and their quantum
numbers see, for instance G.~B. West, 
Nucl. Phys. Proc. Suppl., {\bf 54} A, 353 (1997); N.~A.~Tornqvist, ``Summary
of Gluonium95 and Hadron95 Conferences,'' arXiv:hep-ph/9510256;
J.~Terning, arXiv:hep-ph/0204012; R.~C. Brower, 
Int. J. Mod. Phys., A16S1C, 1005 (2001); M. Suzuki, Phys. Rev., D {\bf 65}, 097507 (2002); M.~Teper, 
Nucl. Phys. Proc. Suppl., 109, 134 (2002).

\bibitem{Morningstar:1999rf}
  C.~J.~Morningstar and M.~J.~Peardon,
  Phys.\ Rev.\ D {\bf 60}, 034509 (1999)
  [arXiv:hep-lat/9901004].

\bibitem{comment6}
A general study of knotted flux tubes on the lattice has yet to be preformed and would be quite difficult. We thank Tom DeGrand and David Lin for discussions on this point.

\bibitem{glueball refs}   Many of the $f$-states have been discussed as possible glueball states or as states expected to have some glueball content. The favored states have changed over the years, but without definitive clarification. For instance,  a recent paper \cite{Mennessier:2008kk} focuses on the glueball nature of the $f_0(600)$ but also discusses the gluonic content of numerous other $f$-states and contains numerous references.  The history of glueball models is long and complicated, and so we refer the reader to the literature.

\bibitem{Mennessier:2008kk} 
  G.~Mennessier, S.~Narison and W.~Ochs,
  Phys.\ Lett.\ B {\bf 665}, 205 (2008)
  [arXiv:0804.4452 [hep-ph]]. We thank A. Niemi for pointing out this paper.

\bibitem{comment1} 
Confinement is due to monopole condensation in terms of 't Hooft's dual confinement picture, with flux tubes being chromoelectric.
However, since we as yet have no soliton solutions to the full QCD theory, all we require are tightly knotted/linked flux tubes (chromoelectric, chromomagnetic, or even chromodyonic) of uniform cross section, though we prefer to go with conventional interpretation.

\bibitem{Buniy:2002yx}
  R.~V.~Buniy and T.~W.~Kephart,
  Phys.\ Lett.\ B {\bf 576}, 127 (2003)
  [arXiv:hep-ph/0209339].
  

  
\bibitem{Woltier}
L.~Woltier,
Proc. Nat. Acad. Sci., 44, 489 (1958);
H.~K. Moffatt,
 J. Fluid Mech., 159, 117 (1969);
H.~K. Moffatt,
J. Fluid Mech., 159, 359 (1985).

\bibitem{Taylor} J. B. Taylor, Phys. Rev. Lett. 33, 1139 (1974).

\bibitem{tight}
V. Katritch, et al., Nature, 384, 142 (1996); O. Gonzalez and J. H. Maddocks, Proc. Natl. Acad. Sci. USA, 96, 4769 (1999); P. Pieranski and S. Przybyl, Phys. Rev. E 64, 031801 (2001); A. Stasiak, J. Dubochet, V. Katritch
and P. Pieranski,  ``Ideal knots and their relation to the physics of real knots,'' p-1, World Sci. Pub. (Singapore) 1998.

\bibitem{OH} J. O'Hara,  Topology Appl., 48(2), 147 (1992). 

\bibitem{FHW} M. H. Freedman, Z.-X. He, and Z.  Wang,   Ann.  Math., Second Series, 139(1), 1 (1994).

\bibitem{knotenergy}
R. A. Litherland, J. Simon, O. Durumeric, and E. J. Rawdon,  Topology Appl., 91(3), 233 (1999);
O. Gonzalez, and J. H. Maddocks, 
Proc. Natl. Acad. Sci. USA, 96(9), 4769  (1999);
A. Nabutovsky,  Comm. Pure Appl. Math., 48(4), 381  (1995).

\bibitem{RR}
F. Maggioni and R. L. Ricca,
Proc. R. Soc. A 2009 465, 2761 (2009).

\bibitem{CKS}
J. Cantarella, R. B. Kusner, and J. M. Sullivan, Invent.  Math., 150(2), 257  (2002). 

\bibitem{DH} H. Doll and J. Hoste, Math. Computation, 57, 747 (1991). 

\bibitem{HT} J. Hoste and M. B. Thistlethwaite and J. R. Weeks, Math. Intelligencer, 20, 33 (1998).
  
  \bibitem{PP} ``High resolution portrait of the ideal trefoil knot,'' P. Pieranski and S. Przybyl,  preprint, January (2012).

\bibitem{FF+FA}
See \cite{Thacker:2008dr}  for recent results on instanton or the lack thereof in the QCD vacuum \cite{Witten:1978bc} and \cite{Witten:1988hf} for a study of knots and Chern-Simons terms in gauge theories.

\bibitem{Thacker:2008dr}
  H.~B.~Thacker,
  PoS LATTICE 2008, 260 (2008)
  [arXiv:0810.4131 [hep-lat]].

\bibitem{Witten:1978bc}
  E.~Witten,
  Nucl.\ Phys.\ B {\bf 149}, 285 (1979).
  
\bibitem{Witten:1988hf}
  E.~Witten,
  Commun.\ Math.\ Phys.\ {\bf 121}, 351 (1989).


\bibitem{ridgerunner} 
T. Ashton, J. Cantarella, M. Piatek, and E. J. Rawdon,    
Experimental Mathematics, 20(1), 57, (2011).  

 

\bibitem{CLR} 
J. Cantarella, A. LaPointe, E. J. Rawdon, J. Phys. A 45, 1 (2012).

\bibitem{GriffithsEM} D. J. Griffiths, {\em Introduction to Electrodynamics}, 3rd ed.,  Benjamin Cummings Press (1999).

\bibitem{elasticrods} M. Chamekh,  S. Mani-Aouadi and M. Moakher, 
Computer Methods in Applied Mechanics and Engineering, 198(47-48), 3751 (2009);
J. Spillmann and M. Teschner,    
EUROGRAPHICS, 27(2), 497 (2008).



\bibitem{comment3}
To describe chromomagnetic flux tubes, we use the Lagrangian density ${\cL}=\frac{1}{2}\tr F_{ij}F^{ij}-V$, to which we add $\tr\lambda\{\Phi_B/(\pi a^2)-\frac{1}{2}\eps^{ijk}n_iF_{jk}\}$.
Then variation gives $D^j(F_{ij}-\frac{1}{2}\eps_{ijk}n^k\lambda)=0$ with resulting constant fields $F_{ij}=(\Phi_B/\pi a^2)\eps_{ijk}n^k$.
Again, the energy is positive and proportional to $l$.

\bibitem{flux} 
In a discussion of flux quantization in non-Abelian gauge theories, it is important to keep the following fact in mind.
For a static configuration, and using the temporal gauge, we can avoid the creation of chromomagnetic flux lines inside the flux tubes~\cite{Nielsen-Olesen}.
A dual argument shows chromoelectric flux creation can be avoided as well.

\bibitem{soliton}
For other approaches to solitonic bags see:
R.~Friedberg and T.~D. Lee,
Phys. Rev., D {\bf 18}, 2623 (1978);
R.~Goldflam and L.~Wilets,
Phys. Rev., D {\bf 25}, 1951 (1982);
G.~Clement and J.~Stern,
Phys. Rev., D {\bf 34}, 1581 (1986).

\bibitem{MIT-bag}
A.~Chodos, R.~L. Jaffe, K.~Johnson, C.~B. Thorn, and V.~F. Weisskopf,
 Phys. Rev., D {\bf 9}, 3471 (1974); T.~DeGrand, R.~L. Jaffe,
K.~Johnson, and J.~E. Kiskis,
 Phys. Rev., D {\bf 12}, 2060 (1975);
A.~Chodos, R.~L. Jaffe, K.~Johnson, C.~B. Thorn, and V.~F. Weisskopf,
Phys. Rev., D {\bf 9}, 3471 (1974);
T.~DeGrand, R.~L. Jaffe, K.~Johnson, and J.~E. Kiskis,
Phys. Rev., D {\bf 12}, 2060 (1975).

\bibitem{Breakstone:1990at}
  A.~Breakstone, {\it et al.},
  Z.\ Phys.\ C {\bf 48}, 569 (1990).

\bibitem{Rolfsen}
D.~Rolfsen,
\newblock {\em Knots and Links},
\newblock Publish or Perish, 1990.

\bibitem{Kauffman}
L.~H. Kauffman,
\newblock {\em Knots and physics},
\newblock World Scientific, 2001.

\bibitem{reconnection}
E.~Priest and T.~Forbes,
\newblock {\em Magnetic Reconnection: {MHD} Theory and Applications},
\newblock Cambridge, 2000.

\bibitem{Casher:1978wy} 
  A.~Casher, H.~Neuberger and S.~Nussinov,
  Phys.\ Rev.\ D {\bf 20}, 179 (1979).



\bibitem{Neuberger:1979tb} 
  H.~Neuberger,
  Phys.\ Rev.\ D {\bf 20}, 2936 (1979).

\bibitem{Casher:1979gw} 
  A.~Casher, H.~Neuberger and S.~Nussinov,
  Phys.\ Rev.\ D {\bf 21}, 1966 (1980).

\bibitem{DegrandBook}
For a discussion see ``Lattice Methods for Quantum Chromodynamics,'' T. Degrand and C. DeTar,
World Scientific Publishing Company (2006), p96.

\bibitem{commentfit}
We note that attempts have been made to fit many of the $f_J^{++}$ states into the quark model.

 

\bibitem{KnotSym} For a full discussion of these symmetries see M. Berglund, et al., arXiv:1010.3234.

\bibitem{Spheromaks}
P.~M.~Bellan, {\it Spheromaks: a practical application of magnetohydrodynamic dynamos and plasma self-organization}, Imperial College Press (London) 2000. 

\bibitem{Oppermann:2010uy}
  See N.~Oppermann, H.~Junklewitz, G.~Robbers and T.~A.~En{\ss}lin,
  arXiv:1008.1246 [astro-ph.IM], and references therein.

\bibitem{lightning}
A.~F. Ra\~{n}ada, M.~Soler, and J.~L. Trueba,
Phys. Rev., E {\bf 62}, 7181 (2000).

\bibitem{Bose-Einstein}
Y.~M.~Cho,
arXiv:cond-mat/0112325.
  For recent work and discussion of knot-like solutions in BECs and other systems see Y.~Kawaguchi, M.~Nitta and M.~Ueda,
  Phys.\ Rev.\ Lett.\ {\bf 100}, 180403 (2008)
  [Erratum-ibid.\ {\bf 101}, 029902 (2008)]
  [arXiv:0802.1968 [cond-mat.other]] and references therein.
  Older references can be found in \cite{Buniy:2002yx}. For instance, Faddeev and Niemi, 
  [Nature 387, 58 (1997)] find a knot solutions in field theory and identify 
  their knot with the the $\eta(1440)$,
  L.~Faddeev, A.~J.~Niemi and U.~Wiedner,
  Phys.\ Rev.\ D {\bf 70}, 114033 (2004).
   
\bibitem{Buniy:2004zy}
  R.~V.~Buniy and T.~W.~Kephart,
  Proc. Coral Gables 2003, published in
  Int.\ J.\ Mod.\ Phys.\ A {\bf 20}, 1252 (2005)
  [arXiv:hep-ph/0408027].

\bibitem{Buniy:2004zw}
  R.~V.~Buniy and T.~W.~Kephart,
{\it Physical and Numerical Models in Knot Theory Including Applications to the Life Sciences}, ed. J. Calvo, K. C. Millett, E. J. Rawdon and A. Stasiak. World Scientific, 2005. p. 45. 
  arXiv:hep-ph/0408025.





\end{thebibliography}
\end{document}